\documentclass[journal]{IEEEtran}

\usepackage{lineno}

\usepackage{framed,multirow}
\usepackage{amsmath,graphicx, subfigure,  epstopdf, units, pifont, enumerate, setspace, mathtools, amsthm, array, todonotes, algorithmic, amssymb, latexsym}

\usepackage{url}

\usepackage{xcolor}
\definecolor{newcolor}{rgb}{.8,.349,.1}

\DeclareMathOperator{\rank}{rank}
\DeclareMathOperator*{\argmin}{\arg\!\min}

\begin{document}

\title{PrecoG: an efficient unitary split preconditioner for the transform-domain LMS filter via graph Laplacian regularization}

\author{Tamal Batabyal,
        Daniel Weller, Jaideep Kapur, Scott T. Acton~\IEEEmembership{Fellow,~IEEE,}}

\maketitle
\begin{abstract}
Transform-domain least mean squares (LMS) adaptive filters encompass the class of algorithms where the input data are subjected to a data-independent unitary transform followed by a power normalization stage as preprocessing steps. Because conventional transformations are not data-dependent, this preconditioning procedure was shown theoretically to improve the convergence of the LMS filter only for certain classes of input data. However, in reality if the class of input data is not known beforehand, it is difficult to decide which transformation to use. Thus, there is a need to devise a learning framework to obtain such a preconditioning transformation using input data prior to applying on the input data. It is hypothesized that the underlying topology of the data affects the selection of the transformation. With the input modeled as a weighted graph that mimics neuronal interactions, PrecoG obtains the desired transform by recursive estimation of the graph Laplacian matrix. Additionally, we show the efficacy of the transform as a generalized split preconditioner on a linear system of equations  and in Hebb-LMS settings. In terms of the improvement of the condition number after applying the transformation, PrecoG performs significantly better than the existing state-of-the-art techniques that involve unitary and non-unitary transforms. 
\end{abstract}



%


\section{Introduction}
\label{intro}
In 1960, Bernard Widrow and Ted Hoff~\cite{widrow1960adaptive} proposed a class of \textit{least mean squares} (LMS) algorithms to recursively compute the coefficients of an N-tap finite impulse response (FIR) filter that minimizes the output error signal. This computation is achieved by a stochastic gradient descent approach where the filter coefficients are evaluated as a function of the \textit{current error} at the output. 
The LMS algorithm and its variants were subsequently used in myriads of applications, including echo cancellation~\cite{farhang1997fast, ma2010adaptive}, inverse modeling~\cite{widrow1978adaptive}, system identification, signal filtering~\cite{chen2000ecg, he2004removal} and several others.

Two of the major issues with this approach are the convergence speed and stability. The filter coefficients (or weights) converge in mean while showing small fluctuation in magnitude around the optimal value. The convergence speed depends on the condition number of the autocorrelation matrix of the input, where a condition number close to unity connotes a fast and stable convergence. Later, adaptive algorithms, such as LSL (least square lattice) and GAL (gradient adaptive lattice)~\cite{paleologu2008gradient, fan1993gal} filters were designed to achieve faster convergence, immunity to poor condition number of input autocorrelation matrix, and better finite precision implementation compared to the LMS filter. However, these stochastic gradient filters may sometimes produce significant numerical errors, and the convergence is poor compared to recursive least squares (RLS) filters~\cite{Haykin:1996:AFT:230061}. Due to the fact that the nature of the autocorrelation matrix is data-dependent, improving its condition number by using a transformation is a way to circumvent the convergence issue in the case of real time data.

In order to obtain well-conditioned autocorrelation matrix of any real world input data, we transform the input \textit{a priori}, which is popularly known as transform-domain LMS (TDLMS) (\textbf{See Appendix for the LMS and TDLMS algorithms}). The discrete Fourier transform (DFT), discrete cosine transform (DCT)~\cite{zhao2009stability, hosur1997wavelet, costa2001stochastic, kim2000performance} 
\begin{figure*}[!t]
\vspace*{-0.5cm}
	\centering
	\subfigure[]{\includegraphics[width=11 cm,height=15cm]{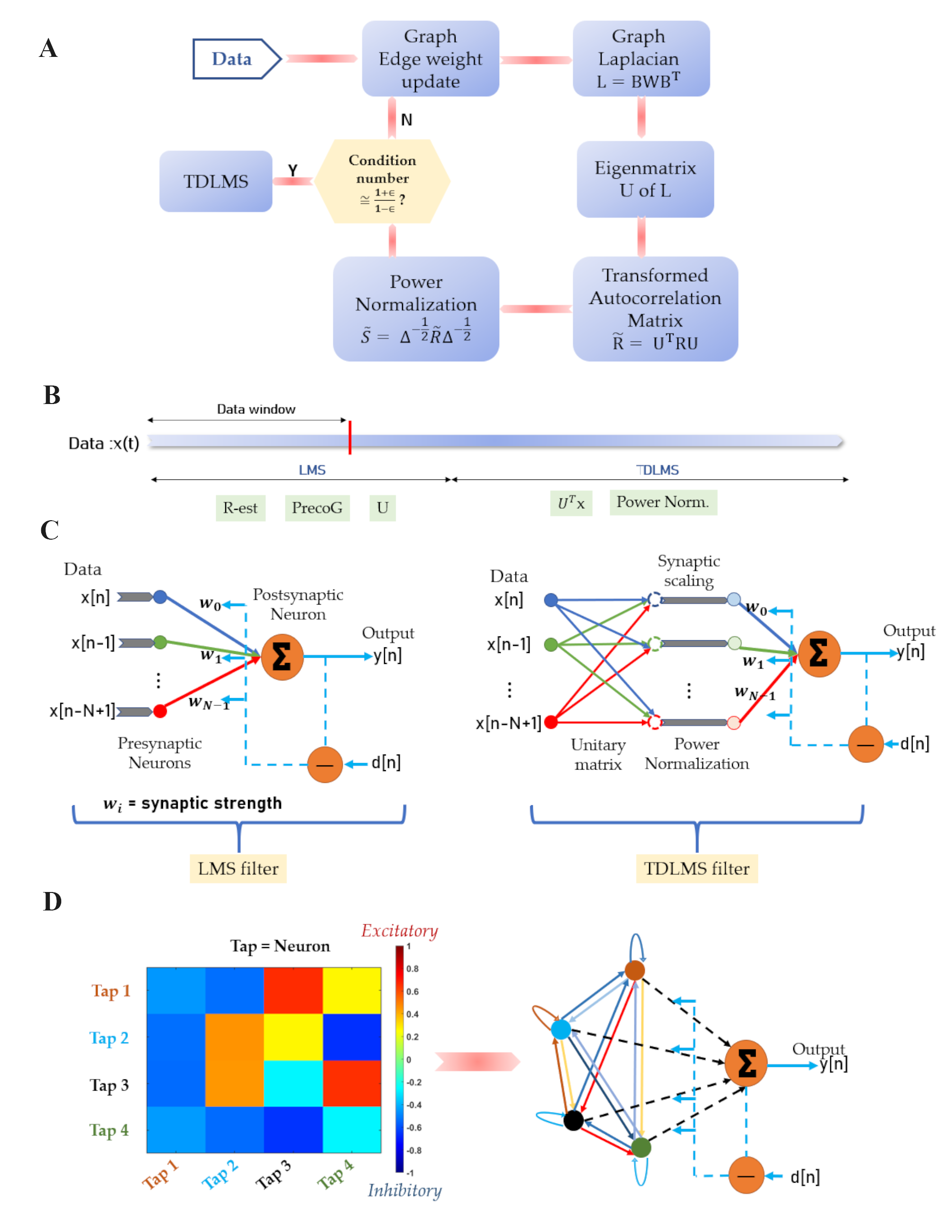}}	
	\caption{(A) A schematic of our algorithm about how the PrecoG transformation is computed. (B) Online mode of LMS in case of PrecoG. Here, the autocorrelation is estimated using an initial window of data. During that time, classical LMS is executed. Once the U is computed, the rest of the data is channeled in TDLMS mode. (C) Equivalence between biological neurons and its simplified computational analogue in LMS. Synaptic scaling is achieved via power normalization stage. (D) A preconditioning matrix U is shown. Its use in the TDLMS architecture is shown in color.}
	\label{Algflow}
\vspace*{-.5cm}
\end{figure*}

and others act as suitable off-the-shelf transformations of the input data for such problems. The aforementioned step is immediately followed by a power normalization stage~\cite{farhang1998transform, beaufays1995transform} and then used as input to the LMS filter. As a geometrical interpretation, the unitary transformation rotates the mean square error (MSE) hyperellipsoid without changing its shape on the axes of LMS filter weights~\cite{beaufays1995transform}. The rotation tries to align the axes of the hyperellipsoid to the axes of weights.  
The power normalization is crucial in enhancing the speed of convergence of LMS filter. The normalization forces the hyperellipsoid to cross all the axes at equal distance from the center of the hyperellipsoid. For a perfect alignment after the transformation, the normalization step turns the MSE hyperellipsoid into a hypersphere~\cite{beaufays1995transform}. It is important to note that these off-the-shelf transformations are still not data-dependent. They work relatively well for certain classes of data having autocorrelation matrices with special structure (eg. Toeplitz).
 
The TDLMS filter is flexible as it does not attempt to change the working principles and the architecture of LMS filter. Therefore, the transform-domain module can precede other algorithms, such as RLS, GAL and LSL. Notice that the conventional unitary transformations are independent of the underlying data, hence not optimal in regularizing condition numbers of the autocorrelation matrices of arbitrary real-time datasets. As an example, the DCT has been shown to be near-optimal for Toeplitz matrices. However, the DCT loses its near-optimality in conditioning sparse linear systems. 

From a different perspective, the transformation of a matrix, such as autocorrelation matrix to improve the condition number is regarded as a subproblem of preconditioning of matrices~\cite{chen2005matrix,benzi2002preconditioning,cao2011splitting,zheng2017extended}. Jacobi~\cite{dauphin2015equilibrated, chapelle2011improved}, Gauss-Seidel~\cite{dziekonski2010jacobi}, approximate inverse~\cite{tang1999toward, benzi2000robust}, incomplete LU factorization~\cite{chow2015fine} preconditioners are examples of such data-dependent transformations that utilize a decomposition of the input matrix consisting of coefficients of linear equations. These algorithms are well-suited for solving linear system of equations. In short, the transformation that improves the performance of the LMS filter can also be applied to solving linear systems of equations.

However, not all the strategies for solving linear systems of equations using matrix-preconditioner are suitable for TDLMS.
With careful attention, it can be seen that there is a difference between TDLMS and linear systems in terms of the usage of a preconditioner, as explained below. The preconditioning action is \textit{implicit} in TDLMS filters (split preconditioner). The autocorrelation matrix is not explicitly used in the LMS architecture. Instead, it is the input data or the transformed input data (in case, we transform the data) that are channeled through the LMS lattice and the update of weights is based on error-correcting learning. While expressing the mean square error at the output as a function of filter weights, it appears that the transformation matrix at the input is attempting to condition the input autocorrelation matrix and the convergence of the LMS algorithm depends on the input autocorrelation matrix. On the other hand, in case of solving linear system of equations, $(Ax=b)$ type, the use of a preconditioner is \textit{explicit}, which is $M^{-1}Ax = M^{-1}b$, with \textit{M} as a preconditioner matrix, such as the Gauss-Seidel type. 

Let $A$ be a matrix to be preconditioned by another matrix $\zeta$. $\zeta$ is said to be a left, right, and split preconditioner if $\zeta^{-1}A$, $A\zeta^{-1}$, and $U_1^{-1}AU_2^{-T}$ with $\zeta = U_1U_2^T$ respectively provide improved condition numbers compared to that of \textit{A}. Gauss-Seidel, incomplete LU, and approximate inverse are examples of a left preconditioner. The transformations in TDLMS algorithm are the unitary split preconditioner type. By unitary, we have  $\big(U_1U_2^T = I\big)$ and $U_1 = U_2 = U$. Therefore, it is the unitary split preconditioner matrix that can be applied to TDLMS as well as to solving linear systems of equations. In PrecoG, we aim to learn such unitary split preconditioner from input data. 
 In addition, as the transformation is unitary, the energy of input data remains unchanged after transformation. 
 
Unlike the off-the-shelf, data-independent transformations in TDLMS, the derivation of our unitary split conditioner is motivated by the topology of the structured input data. The topology determines neighborhood relationship between data points, which can be represented using graph-theoretic tools~\cite{maretic2017graph,pavez2018learning,rabbat2017inferring}. In recent years, manifold processing and regularization have shown promise in different areas of research~\cite{abernethy2010graph,  hallac2015network}. Based on such evidence, we hypothesize that the intrinsic topology of the input data affects the construction of a suitable preconditioner matrix.   
We estimate a data manifold that provides an alternate set of basis acting as a split preconditioning matrix. The data when projected onto the basis are expected to be decorrelated.

The \emph{main contributions} of this work are as follows.
\begin{itemize}
    \item PrecoG provides an $optimization$ framework that finds the desired unitary transformation for the preconditioning matrix. We iteratively estimate the underlying topology leveraging graph theory, followed by the computation of desired unitary transformation by using the graph Laplacian and first order perturbation theory. 
    PrecoG significantly outperforms all conventional off-the-shelf transformations.
    
    \item We show that our approach is equally applicable in preconditioning arbitrary linear systems apart from ameliorating the convergence of LMS filters.
    
    \item Another advantage of PrecoG is that it can be applied without having a prior knowledge about the process that generates the input data. Estimated autocorrelation matrix serves our purpose.
    
    \item PrecoG shows its efficacy in supervised and unsupervised learning tasks. Using analogy from Neuroscience, PrecoG is shown to work in the Hebb-LMS paradigm.  
\end{itemize}
 To our knowledge, no mathematical framework has been presented so far that could potentially generate such matrices from the data.

\subsection{Why a graph theoretic approach?}
The relevance of using graphs is rooted in neuroscience.
The LMS algorithm has been pivotal in the overall error-correcting learning paradigm, mostly because of its mathematical elegance and simple interpretation. Its inventors, Widrow and Hoff, named this adaptive filter as \textit{Adaline} in 1957. Frank Rosenblatt~\cite{rosenblatt1957perceptron} successfully devised a prototypical version of a supervised binary classifier, called the \textit{Perceptron}, that superficially mimics the action-potential based firing of a single neuron, where the weight update step utilizes the LMS algorithm. Here, the weight is synonymous with the synaptic strength in neuroscience. That analogy is backed up by the facty that the weights are continuously updated real numbers and the final weight values are `endured' (converged), they have a weak notion of \emph{plasticity} at the synapse level which is observed when the network is sufficiently trained. In reality, the plasticity in neurons in our brain is a consequence of immensely intricate molecular process~\cite{abbott2000synaptic, rudy2014neurobiology}, and the plasticity of the adaptive filter weights are extremely simple parallel. The Perceptron was extended to multiple layers to effectively tackle problems having nonlinear decision boundaries (such as XOR function). Subsequently, numerous (if not the majority of) neural network models including convolutional networks have used the LMS algorithm to update the network weights.

Tuning synaptic strengths (filter weights) is useful but of limited scope, as it delivers an incomplete portrait of the actual task - adapting the filter weights with the datastream in this work. Researchers have gathered evidence supporting the conclusion that neuronal level dynamics orchestrate complex behavioral functions~\cite{stam2014modern, josselyn2020memory,bargmann2013connectome, aoi2020prefrontal}. In general, neurons inhibit or excite other neurons through specific neuromodulators spewed at the synaptic clefts. In an LMS filter, the filter taps can be conceived as neurons.
For example, in our LMS filter setting, we have $N$ neurons that conduct signals, which are weighted (tap weights), summed and passed to another neuron. However, contrary to the neurons in our brain, these filter taps are independent of each other in the traditional LMS implementation. PrecoG generalizes the filter by incorporating interactions among the taps. This interaction is a function of the input data.

The question now becomes: how do these filter taps (equivalently neurons) interact (excite/inhibit) with each other \emph{over time} in order to \emph{expedite} the convergence of tap weights (synaptic strengths)?  The graph is an efficient tool to encode such interaction~\cite{rubinov2010complex}. It is worth mentioning a couple of points in this aspect - 1) the dynamics of interaction and the synaptic strength influence each other every time the graph parameters are estimated; 2) we are simply looking for the excitatory/inhibitory drive of each filter tap onto other taps. The resultant graph is simple and undirected. However, a neuron can impart an excitatory or inhibitory drive onto itself  (\textbf{Figure}). In biological neurons, these synapses of self-stimulation are called autapses~\cite{ikeda2006autapses}. These are abundant among the fast-spiking inhibitory neurons in the neocortex.

\vspace*{-0.4cm}
\section{Graph Theory (in brief)}

\label{gs}
A graph can be compactly represented by a triplet $\big(\mathcal{V}, \mathcal{E}, \boldsymbol{w}\big)$, where $\mathcal{V}$ is the set of vertices, $\mathcal{E}$ the set of edges, and $\boldsymbol{w}$ the weights of the edges. For a \emph{finite} graph, $|\mathcal{V}| ~=~ N$ which is a finite positive integer, and $|\bullet|$ is the cardinality of a set. By denoting $w_{ij}\in\boldsymbol{w}$ as a real positive weight between two vertices $i$ and $j$ with $i,j\in\{1,2,...,N\}$, the adjacency matrix, $A$ of $\mathcal{G}$ can be given by $a_{ij} = w_{ij}$ with $a_{ii} = 0$ for a graph with no self-loop. $A\in\mathcal{R}^{N\times N}$ is symmetric for an undirected graph and can be sparse based on the number of edges. The incidence matrix, $B\in\mathcal{R}^{N\times |\mathcal{E}|}$ of $\mathcal{G}$ is defined as $b_{ij} = 1~or~-1$ where the edge $j$ is incident to or emergent from the vertex $i$. Otherwise $b_{ij} = 0$. The graph Laplacian $L\in\mathcal{R}^{N\times N}$, which is a symmetric positive-semidefinite matrix, can be given by $L = BWB^T$, where $W$ is a diagonal matrix containing $\boldsymbol{w}$. Determining the topology of input data refers to the estimation of $A$ or $L$ depending on the formulation of the problem at hand. 

\section{Problem statement}
\vspace*{0.1cm}
\label{pst}
Let $x_k$ = $[x(k)~x(k-1)~\dots~x(k-N+1)]$ be a $N$-length real valued tap-delayed input signal vector at $k^{th}$ instant. The vector representation is convenient for estimating the input autocorrelation as an ergodic process. Let the autocorrelation matrix, denoted by $R_N$ be defined as $R_N = E\big(x_Nx_N^T\big)$. We assume that $\Delta_Y$ is the main diagonal of a square matrix $Y$ $(\Delta_Y = diag(Y))$. Following this notation, after power normalization the autocorrelation matrix becomes $S_N = \Delta_{R_N}^{-\frac{1}{2}}R_N \Delta_{R_N}^{-\frac{1}{2}}$. In general, the condition number of $S_N$, $\chi_{S_N}$, happens to be significantly large in practical datasets. For example, the condition number of the autocorrelation matrix of a Markov process with signal correlation factor as 0.95 has a $\chi$ of $\mathcal{O}(10^3)$.  
Notice that we seek $U_N$ to minimize the condition number of $S_N$. Let a unitary transformation be $U_N$ $(U_NU_N^T = I)$ such that the transformed autocorrelation matrix becomes $\tilde{R}_N = E\big[U_N^Tx_kx_k^TU_N\big]$. Next, $\tilde{R}_N$ is subjected to a power normalization stage that produces $\tilde{S}_N = \Delta_{\tilde{R}_N}^{-\frac{1}{2}}\tilde{R}_N \Delta_{\tilde{R}_N}^{-\frac{1}{2}}$. Precisely, we want the eigenvalues of $\lim_{N\to\infty}\tilde{S}_N\in\big[1-\epsilon_2,~1+\epsilon_1\big]$, where $\epsilon_1$ and $\epsilon_2$ are arbitrary constants such that $\chi_{max}\simeq \frac{1+\epsilon_1}{1-\epsilon_2}$. A schematic of our algorithm is given in Fig.~\ref{Algflow}. 

Let us take an example of a $1^{st}$ order Markov input with the signal correlation factor $\rho$ and autocorrelation matrix, $R_{N}$ as
\begin{eqnarray}
R_{N} = E[\boldsymbol{x}_{k}\boldsymbol{x}_{k}^{H}]=
	\begin{pmatrix}
		1 & \rho & \cdots & \rho^{n-1} \\
		\rho & 1 & \cdots & \rho^{n-2} \\
		\vdots & \vdots & \ddots & \vdots \\
		\rho^{n-1} & \rho^{n-2} & \cdots & 1
	\end{pmatrix}
\label{acr}
\end{eqnarray} 
It is shown in~\cite{beaufays1995transform} that $\chi_{S_{N}}\simeq(\frac{1+\rho}{1-\rho})^2$, which suggests that $\epsilon_1 = \rho^2+2\rho$, and $\epsilon_2 = 2\rho-\rho^2$. After applying the DFT, the condition number becomes $\lim_{N\to\infty}\chi_{\tilde{S}_{N}} = (\frac{1+\rho}{1-\rho})$, which indicates that $\epsilon_1=\epsilon_2=\rho$. On applying the DCT, $\lim_{N\to\infty}\chi_{\tilde{S}_{N}} = 1+\rho$ with $\epsilon_1 = \rho$ and $\epsilon_2 = 0$. 

\vspace*{0.2cm}
\section{Methodology}

\label{method}
The search for $U_N$ is carried out through an iterative optimization of an associated cost function. It can be argued from  section~\ref{pst} that the optimal convergence properties are obtained when $\tilde{S}_{N}$ converges to the identity matrix in the \textit{rank zero perturbation} sense \cite{beaufays1995transform}: $A$ and $B$ with $\eta=A-B$  have the same asymptotic eigenvalue distribution if
\begin{eqnarray}
\lim_{N\longrightarrow\infty}\rank(\eta) = 0.
\vspace*{-0.2cm}
\end{eqnarray}
In our case, with $\lambda$ as an eigenvalue, this translates to 
\begin{equation}
\lim_{N \to \infty}\det(\tilde{S}_{N} - \lambda\mathbb{I}_{N}) = 0,
\vspace*{-0.2cm}
\label{sn}
\end{equation}
 which can be expanded as,
\begin{equation}
\lim_{N \to \infty}\det\left(\Delta_{\tilde{R}_{N}}^{-1/2}\tilde{R}_{N}\Delta_{\tilde{R}_{N}}^{-1/2} - \lambda\mathbb{I}_{N}\right) = 0.
\vspace*{-0.2cm}
\label{rn}
\end{equation}  
Eq.~(\ref{rn}) can be rearranged as,
\begin{equation}
\lim_{N \to \infty}\det\left(\tilde{R}_{N} - \lambda\Delta_{\tilde{R}_{N}}\right) = 0,
\vspace*{-0.2cm}
\label{costfn1}
\end{equation}  
which is a quadratic polynomial of $U_N$ as $\tilde{R}_N = U_N^TR_NU_N$.
Given the orthonormality of the eigenvectors $U_{n}$, we can rewrite (\ref{costfn1}) as
\begin{equation}
U_{N}=\argmin_{U_{N}\in O(N)}\left[\det\left(\tilde{R}_{N} - \lambda\Delta_{\tilde{R}_{N}}\right)\right].
\vspace*{-0.2cm}
\label{costfn2}
\end{equation}
Here, $O(N)$ is the set of unitary matrices. However, in presence of the determinant in (\ref{costfn2}), obtaining a closed-form expression for $U_{N}$ is difficult to obtain. To overcome this obstacle, we apply the Frobenius norm in (\ref{fincost}). 
\begin{equation}
U_{N}=\argmin_{U_{N}\in O(N)}\|\tilde{R}_{N} - \lambda\Delta_{\tilde{R}_{N}}\|_{F}^{2}.
\vspace*{-0.2cm}
\label{fincost}
\end{equation} 
The Frobenius norm imposes a stronger constraint compared to (\ref{sn}). In fact, while (\ref{costfn1}) can be solved if at least one column of $\tilde{R}_N - \lambda\Delta_{\tilde{R}_{N}}$ can be expressed as a linear combination of rest of the columns, (\ref{fincost}) becomes zero only when $\tilde{R}_N - \lambda\Delta_{\tilde{R}_{N}}$ is a zero matrix.  In effect, it reduces the search space of $U_N$. It is due to the fact that the set of $U_N$ that solves eq.~(\ref{fincost}) is a subset of the $U_N$ that are also the solutions of eq.~(\ref{costfn2}). 
\vspace*{-0.05cm}
Here, we address two aspects of the problem. First, (\ref{fincost}) attempts to minimize the difference between $\tilde{R}_N$, which is $U_N^TR_NU_N$, and the scaled diagonal matrix of $\tilde{R}_N$. This is necessary because it accounts for the spectral leakage~\cite{beaufays1995transform} as mentioned later in this section. The second aspect is that (\ref{fincost}) seems to diagonalize $\tilde{R}_N$ apart from attempting to make the eigenvalues unity only. Here, a solution may be hard to obtain in practice. So, we relax the unity constraint by forcing the eigenvalues to lie within a range $[1-\epsilon_2,~1+\epsilon_1]$. By enforcing the constraint, (\ref{fincost}) with parameters $p ~=~ (\boldsymbol{w}, \epsilon_{1}, \epsilon_{2})$ becomes
\begin{eqnarray}
U_{N} &=& \argmin_{U_N\in O(N)}|| \underbrace{\tilde{R}_{N} - s_+\Delta_{\tilde{R}_{N}}\|_{F}^{2} +
|| \tilde{R}_{N} - s_-\Delta_{\tilde{R}_{N}}\|_{F}^{2}}_{E(p)}. 
\label{finalCost}
\end{eqnarray}
where $s_+ = 1+\epsilon_{1}$ and $s_- = 1-\epsilon_{2}$ are the upper and lower bounds respectively for the eigenvalues of $S_{N}$. 
Using the Hadamard product notation, we can express $\Delta_{\tilde{R}_{N}}$ as $U_{N}^{T}R_{N}U_{N}\circ\mathbb{I}_{N}$. 
The first two constraints in~(\ref{finalCost}) provide a valley in the space spanned by the eigenvectors in $U_N$ if $R_N$ is positive definite. The valley exists between two surfaces $(1-\epsilon_2)U_N^TR_NU_N\circ I$ and $(1+\epsilon_1)U_N^TR_NU_N\circ I$. At this point, there might be infinitely-many possible solutions. We add a regularizer on $\boldsymbol{w}$ to obtain an acceptable set of solutions.

The undesired result of the imposed restriction given above is that the convergence time may be significant, and the set of solutions of (\ref{finalCost}) in terms of $U_N$ is significantly smaller than that of (\ref{fincost}). Upon approaching a minimum of (\ref{fincost}), the speed of gradient descent algorithm drops significantly. Although it is theoretically expected that $\lim_{N\to\infty}\tilde{S}_N \in [1-\epsilon_2,~1+\epsilon_1]$, in practice, it is difficult to guarantee after a prescribed number of iterative steps.   

\vspace{-0.2cm}
\section{Laplacian parametrization}
\vspace{-0.1cm}
\label{lapa}
The problem of finding a sub-optimal transform by optimizing eq.~(\ref{finalCost}) is solved by leveraging the graph framework. In this framework, the input data is modeled with a finite, single-connected, simple, and undirected graph endowed with a set of vertices, edges and edge weights. For example, for an LMS filter with $N$ taps, an input signal vector $x_k$ has length $N$, which can be represented with $N$ vertices. Basically, each vertex corresponds to one tap of the LMS filter. Using the graph, the unknowns of the optimization in (\ref{finalCost}) are the number of edges and the associated edge weights. A fully-connected graph with N vertices contains $\frac{N(N-1)}{2}$ edges. A deleted edge can be represented with zero edge weight. We denote the the set of unknown parameters as $\boldsymbol{w}$, which is the set of nonzero weights of the graph. 

To find the desired transformation $U_N$, the algorithm initializes the weights $\boldsymbol{w}$ with random numbers sampled from a Gaussian distribution with zero mean and unit variance. Let $W$ is the diagonal matrix containing $\boldsymbol{w}$. Then by definition, the graph Laplacian, which is symmetric and positive semidefinite by construction, is given by $L = BWB^T$. $B$ is the incidence matrix as mentioned in section~\ref{gs}. The spectral decomposition of $L$ provides the matrix of eigenvectors $U$. Finally, $U^Tx_k$ is the transformation that is expected to decorrelate the dataset, which may not be possible due to random initialization. Then the cost function (\ref{finalCost}) helps update the weights and the search for the desired transformation continues in an iterative fashion until the objective conditions are met.

The cost function in eq.~(\ref{finalCost}) is nonconvex. Therefore, the solution is not guaranteed to be a global optimum. In our work, the required solution is obtained through gradient descent with $\mu$ as the step size parameter. 
From section~\ref{gs}, we obtain that $L = BWB^T$. Let $\Theta_i = \frac{\partial L}{\partial w_{i}}$, which can be evaluated as $\frac{\partial L}{\partial w_{i}} = B\frac{\partial W}{\partial w_{i}}B^T$. Using $\mu$ and $\Theta_i$, the update equation is given by 
\begin{eqnarray}
\label{grades}
w_{i}^{t+1} = w_{i}^{t} - \mu Tr\Big(\big[\frac{\partial E(p)}{\partial U_{N}}\big]^T\frac{\partial U_{N}}{\partial w_{i}}\Big); 0<\mu<1 .
\label{gd}
\end{eqnarray}
Here, the computation of $\frac{\partial U_{N}}{\partial w_{i}}$ is perfromed by,
\begin{eqnarray}
\label{grades1}
\frac{\partial u_{k,l}}{\partial w_{i}}~=~Tr\Big(\frac{\partial u_{k,l}}{\partial L}\Theta_i\Big)~= ~Tr\Big(\big[\frac{\partial L}{\partial u_{k,l}}\big]^{-T}\Theta_i\Big),
\end{eqnarray}
where by using $J^{mn} = \delta_{mk}\delta_{nl}$, $\frac{\partial L}{\partial u_{kl}}$ can be given by,
\begin{eqnarray}
L = U_N\Gamma U_N^T \Longrightarrow \frac{\partial L}{\partial u_{kl}} = U_N\Gamma J^{mn} + J^{nm}\Gamma U_N^T.
\label{jmn}
\end{eqnarray}
In eq.~(\ref{grades}), $t$ is the iteration index, and the computation of $\frac{\partial E(p)}{\partial U_{N}}$ is given in the Appendix. 
To prevent each $w_{i}$ from erratic values during iteration, we impose $2-$norm on the weight vector $\boldsymbol{w}$. On adding the regularization term to eq.~(\ref{finalCost}), the new cost function becomes 
\begin{eqnarray}
E_N(\boldsymbol{w},\epsilon_1, \epsilon_2, \beta) = E(p) + \beta(\boldsymbol{w}^T\boldsymbol{w}-1).
\label{haha}
\end{eqnarray}
On differentiating $E_N$ with respect to $\boldsymbol{w}$, we get
\begin{eqnarray}
\label{pde1}
\frac{\partial E_N}{\partial\boldsymbol{w}} = \frac{\partial E(p)}{\partial\boldsymbol{w}} + 2\beta\boldsymbol{w}
\end{eqnarray}
Following eq.~(\ref{pde1}), the iterative update of each weight can be given by,
\begin{eqnarray}
\label{finleqq}
w_i^{t+1} = w_i^t\big(1-2\beta\big)-\mu Tr\Big(\big[\frac{\partial E(p)}{\partial U_{N}}\big]^T\frac{\partial U_{N}}{\partial w_{i}}\Big).
\end{eqnarray}

However, it is difficult to compute the eq.~\ref{gd} because $\frac{\partial L}{\partial u_{kl}}$ is non-invertible. The proof is given in Appendix. One can use diagonal compensation by adding $\alpha I (\alpha < 0.001)$ to $\frac{\partial L}{\partial u_{kl}}$. However, it produces erroneous solution. 

We approach the problem of computing $\frac{\partial U_{N}}{\partial w_{i}}$ using first order eigenvector perturbation. 
We use the result (see Appendix) that if A is a positive-(semi)definite, symmetric matrix and analytic with respect to its entries, then 
\begin{eqnarray}
\label{kato1}
\frac{\partial \boldsymbol{u}_i}{\partial a_{mn}} = \sum_{j \neq i}\frac{1}{(\lambda_i-\lambda_p)}\big<\frac{\partial A}{\partial a_{mn}},\boldsymbol{u}_p\big>\boldsymbol{u}_p;~ \lambda_i\neq\lambda_p.
\end{eqnarray}
Here, $\boldsymbol{u}_i$ and $\lambda_i$ are the $i^{th}$ eigenvector and eigenvalue of $A$ respectively. By definition, the graph Laplacian is symmetric and positive semidefinite. Using eq.~\ref{kato1}, it can be derived that 
\begin{eqnarray}
\label{kato2}
\frac{\partial \boldsymbol{u}_i}{\partial w_j} &=& \sum_{\substack{p \neq i \\
\lambda_p\neq\lambda_i}
}\frac{1-\delta(\lambda_i,\lambda_p)}{(\lambda_i-\lambda_p)}\big<\boldsymbol{u}_i,B\frac{\partial W}{\partial w_{j}}B^T\boldsymbol{u}_p\big>\boldsymbol{u}_p \nonumber\\
&& + \sum_{\substack{q\neq i \\ 
\lambda_q=\lambda_i\neq 0}}
\frac{\delta(\lambda_i,\lambda_q)}{\lambda_i}\big<\boldsymbol{u}_i,B\frac{\partial W}{\partial w_{j}}B^T\boldsymbol{u}_q\big>\boldsymbol{u}_i.\nonumber\\
&& +\sum_{\substack{q\neq i \\ 
\lambda_q=\lambda_i= 0}}
\delta(\lambda_i,\lambda_q)\big<\boldsymbol{u}_i,B\frac{\partial W}{\partial w_{j}}B^T\boldsymbol{u}_q\big>\boldsymbol{u}_i.
\end{eqnarray}
The above formulation can be plugged in $\frac{\partial U_{N}}{\partial w_{i}} = \big[\frac{\partial\boldsymbol{u}_1}{\partial w_j}~\frac{\partial\boldsymbol{u}_2}{\partial w_j}~...~\frac{\partial\boldsymbol{u}_N}{\partial w_j}\big]$ to be used in eq.~\ref{finleqq}.
In eq.~\ref{kato2}, it is evident that the space of incremental changes in an eigenvector with respect to edge weight $w_i$ is a spanned by the eigenvectors $\boldsymbol{u}_i$.  

The data-dependent transformation matrix, $U$ utilizes the autocorrelation matrix of the input data ($\tilde{R}_N$ inside $E(p)$). 
A pertinent question is: how do we get the true autocorrelation matrix for streaming real time data?  In response to that question, we want to remind the reader that with the data being channeled to the filter weights, we simultaneously estimate the input autocorrelation matrix using a data window. This estimated matrix is used in deriving the data-dependent transformation, which is applied on the subsequent data. We show in experiments that this procedure significantly improves convergence.

   
\vspace*{-0.2cm}
\section{Results} 
\vspace*{-0.1cm}
We split the result section in two parts. The first part considers systems where the matrix that is to be conditioned is known a priori. The second part presents the performance of PrecoG on simulated on-line data, where autocorrelation matrices are estimated using a window.

\subsection*{Part-I (Linear systems)}
We show the effectiveness of our approach in preconditioning different matrices against the preconditioners - DCT, DFT, Jacobi (tridiagonal matrix type), GS ($Gauss-Seidel$) and incomplete LU factorization.
To represent the strength of an individual algorithm, we incorporate 
the condition number of each unconditional matrix with the aforementioned methods.  
In order to scale the condition numbers obtained from several methods with respect to ours, we define a metric, \emph{condition ratio} = $\frac{\text{condition number obtained from a method}}{\text{condition number obtained by PrecoG}}$. In some of the results, we compute $log_{10}(\text{condition ratio})$ to mitigate the enormous variance present in the $\text{condition ratio}$ scores.

\begin{figure}[t]
\vspace*{-0.3cm}
	\centering
	\subfigure[]{\includegraphics[width=6.0cm,height=3.6cm]{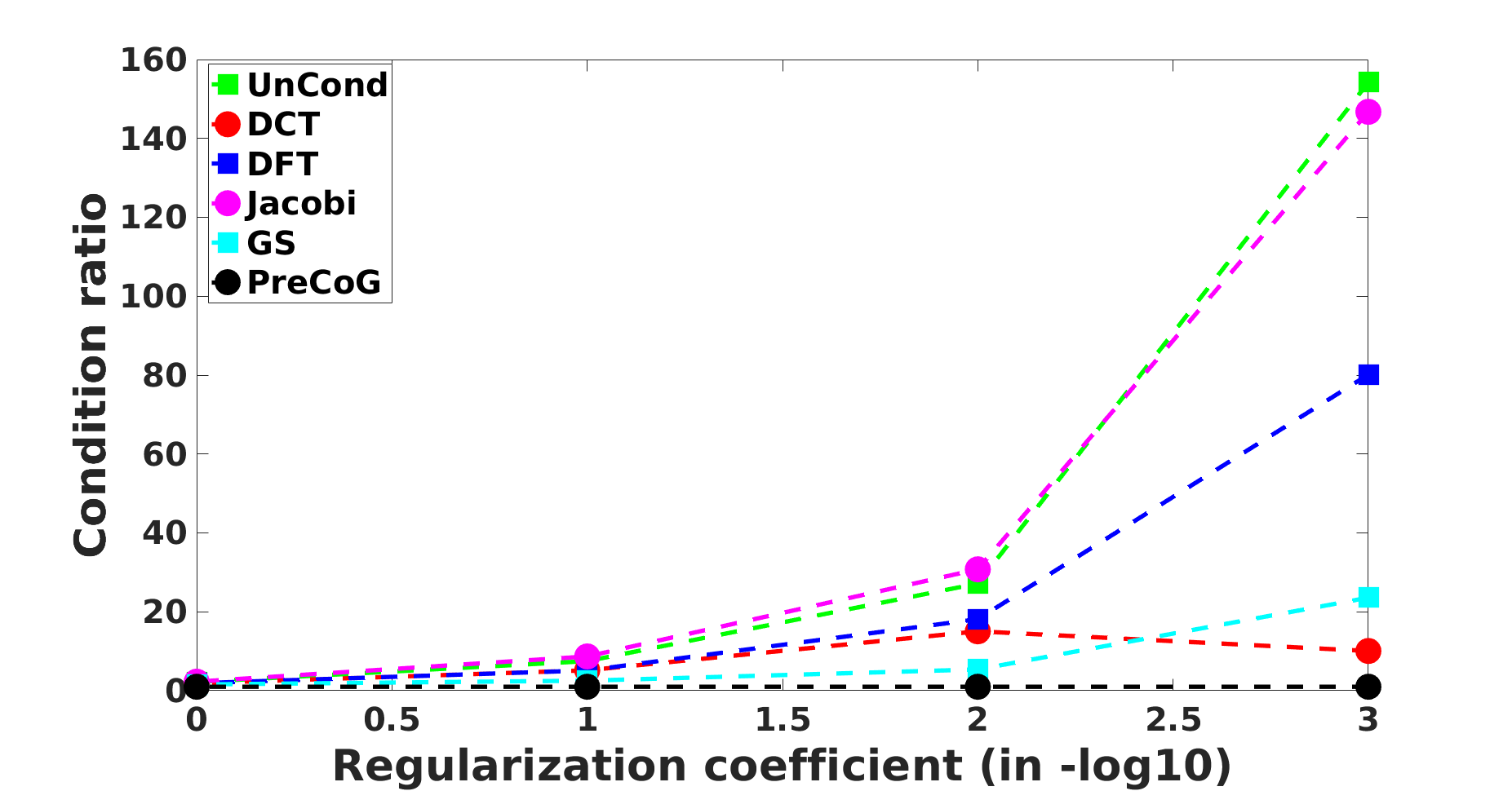}}	
	\subfigure[]{\includegraphics[width=6.0cm,height=3.6cm]{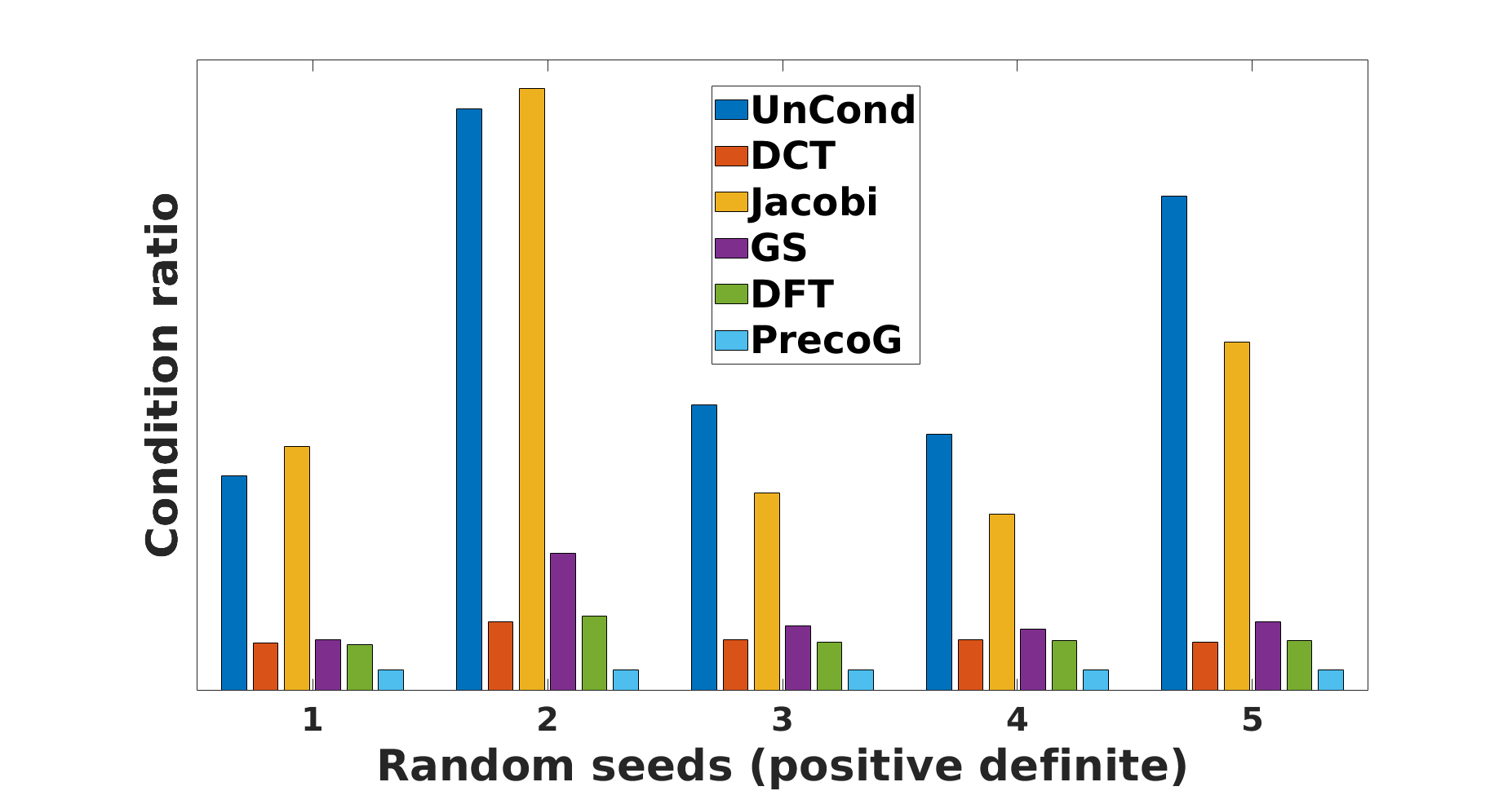}}
	\subfigure[]{\includegraphics[width=6.0cm,height=3.6cm]{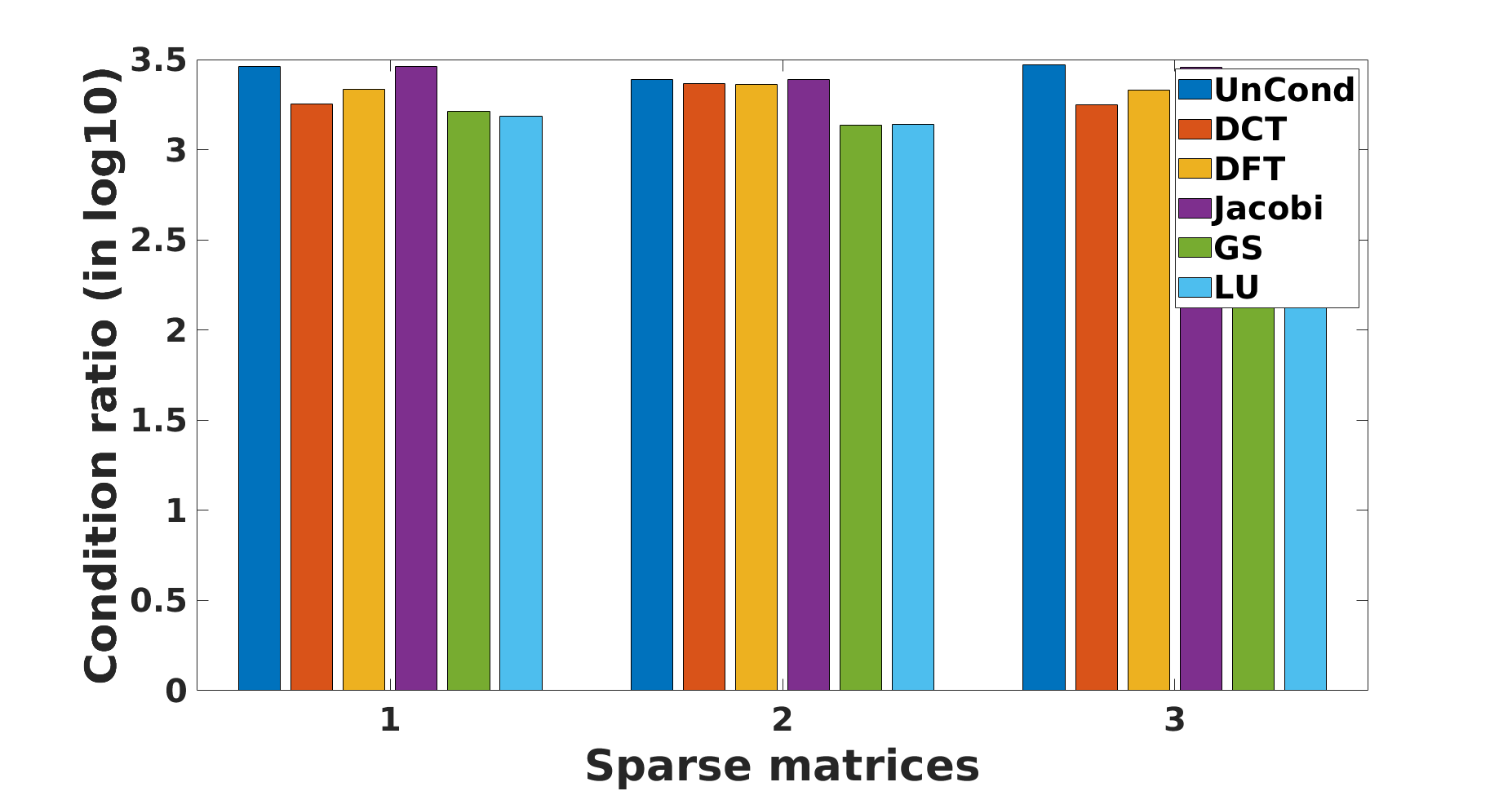}}    
	\vspace*{-0.3cm}
	\caption{\small Condition ratios obtained by applying the algorithms on (a) regularized Hilbert matrices with varied regularization parameters, (b) a set of random matrices containing entries $\sim Gaussian(0,1)$, and (d) random matrices of varied sparsities (for sparse linear systems).}
	\label{figRes}
	\vspace*{-0.4cm}
\end{figure}

First, we apply our method to precondition a Hilbert matrix~\cite{choi1983tricks} which is severely ill-conditioned. Hilbert matrix, $H$ is defined as $H(i,j) = \frac{1}{i+j-1}$. 
 In the experiment, we add a regularizer using $\big(\alpha\mathcal{I}\big)$ with $0<\alpha\le 1$ as the regularization coefficient. The condition ratios of the existing algorithms including PrecoG on preconditioning the Hilbert matrices, which are regularized by changing the $\alpha$, are shown in Fig.~\ref{figRes}(a). Notice that the $X$-axis is given in $-log_{10}$ scale. Therefore, smaller values at $X$ coordinate indicates higher regularization of the Hilbert matrix.  On decreasing the value of $\alpha$, the Hilbert matrix becomes severely ill-conditioned, and the performance of the competitive algorithms except $Gauss-Seidel$ exhibit inconsistent behavior. The DCT performs better near $\alpha = 1~(-log10(\alpha) = 0)$ because of the diagonally dominant nature of the matrix. PrecoG outperformed all the comparative methods. 

We also evaluate our algorithm on five different random positive definite matrices with the values taken from a zero-mean and unit-variance Gaussian process. We regularize the matrices to ensure positive-definiteness. It is evident from Fig.~\ref{figRes}(b), the condition ratios obtained by applying PrecoG outperformed the DCT, Gauss-Seidel, DFT, and Jacobi transformations.  

The condition ratios (in $log10$ scale) with respect to PrecoG on sparse systems of equations are shown in Fig.~\ref{figRes}(c). For PrecoG, $log10$ of the condition ratio is zero. So, it is not shown in the plot. 
The three sparse matrices are random by construction with sparsity $\big(\frac{\text{number of nonzero elements}}{\text{total number of elements}}\big)$ levels as $[\frac{1}{4}, \frac{2}{7}, \frac{1}{5}]$ respectively. PrecoG significantly outperformed the conventional transformations.

\subsection*{Part-II (Simulated process)}
PrecoG is evaluated on two simulated datasets from two different processes - $1^{st}$ order Markov process and $2^{nd}$ order autoregressive process. The LMS algorithm is supervised on these datasets in a sense that the desired output is known beforehand for each dataset. The data is channeled to the filter in on-line mode, and PrecoG is blinded against the customization of the data. The autocorrelation matrices of both processes are Toeplitz. So far, the DCT is known as the near-optimal preconditioner for $1^{st}$ and $2^{nd}$ order Markov processes. We have also considered the performance of PrecoG on Hebbian-LMS learning~\cite{widrow2019nature}. This is an unsupervised form of learning, which is claimed to emulate neuronal learning paradigm.

\subsection{$1^{st}$ order Markov or AR(1) process}
We simulate a first order autoregressive process with a set of signal correlation factors,$\rho$. As mentioned in the third claim of our main contributions, PrecoG is shown to performed significantly well without the prior knowledge of asymptotic autocorrelation matrix of the input data.
\begin{figure}[ht]
\vspace*{-0.3cm}
	\centering
	\subfigure[]{\includegraphics[width=8.0cm,height=4.5cm]{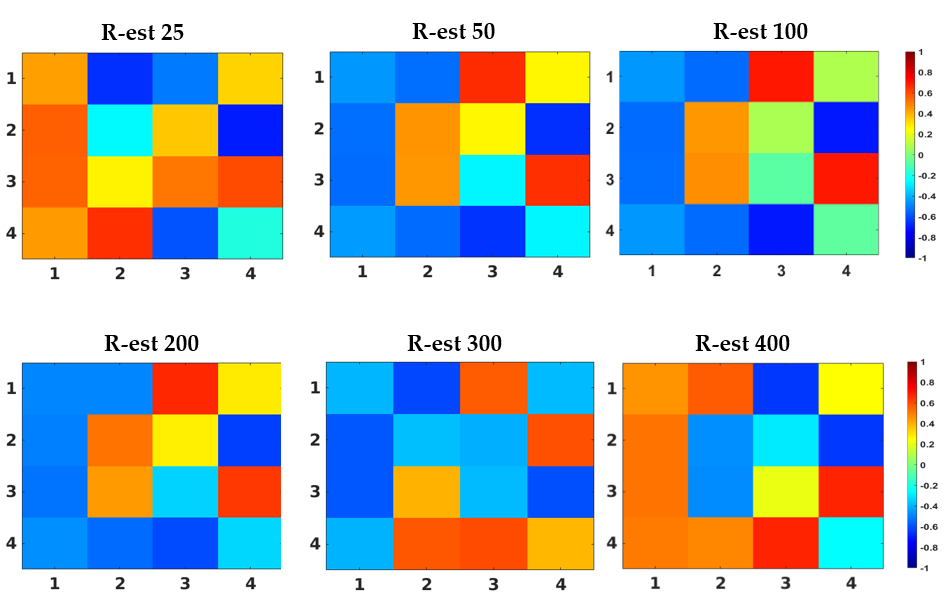}}	
	\subfigure[]{\includegraphics[width=8.0cm,height=2.5cm]{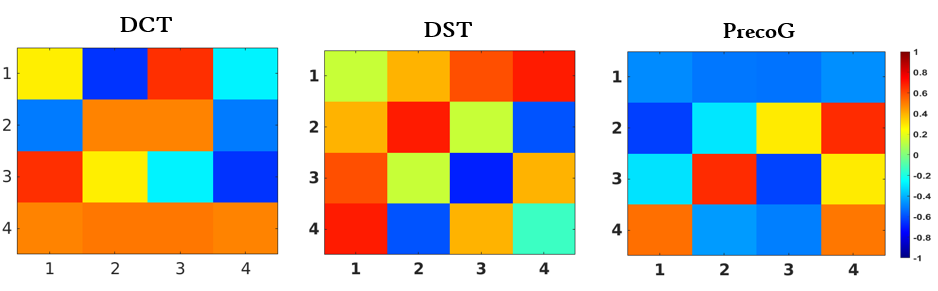}}
	\vspace*{-0.3cm}
	\caption{\small (a) Figure shows the preconditioners which are estimated at different lengths of initial data window. The excitatory and inhibitory drives by each filter tap (a neuron) onto its neighboring taps are clearly visible. PrecoG has considerable magnitude of inhibitory drives among $4$ neurons in the problem. (b) Figure shows the preconditioners - DCT, DST and PrecoG ($125$ window length, AR(1) with $\rho=0.9$).    }
	\label{ar1fig}
	\vspace*{-0.4cm}
\end{figure}

For each $\rho$, we have simulated $2000$ samples of 1D data and convolved the data with a filter defined by coefficients $h = [1~ -0.8~ 6~ 3]$ to generate the desired output. The data power is taken as unity. The output is added with a white Gaussian noise of signal power $0.01$. The goal is to converge the filter weight to the impulse function of the convolution filter. Conventional transformations, such as the DCT and the DST are applied at the beginning. However, due to data-dependent nature of PrecoG, we consider a part of initial data to estimate the time averaged autocorrelation function. The final unitary preconditioner of PrecoG depends on the length of the initial data. The step length of the LMS algorithm is set as $0.002$. For a fixed length of initial data, we have created a search space of L2 coefficients and PrecoG learning rate, to find out the transformation. 

It is evident from Fig~\ref{ar1fig}(a)that the number of taps in the LMS filter is $4$. We have investigated the relationship between these $4$ taps (neurons) by varying the initial data window. Here, $R-est~ 25$ means that the estimated autocorrelation is averaged over the first $25$ data and the `colored' matrix is the transformation. The color describes the strength of excitatory (red) and inhibitory (blue) drives. The relationship (matrix) is asymmetric. As an example to interpret this asymmetricity, it can be seen that for $R-est ~25$, neuron 1 inhibits neuron 2, but neuron 2 applies excitatory postsynaptic potential to neuron 1. The diagonal entries of each transformation are non-empty, indicating the presence of autapses. Fig.~\ref{ar1fig}(b) exhibits the transformation of the DCT, the DST and PrecoG, where the PrecoG transformation is obtained from a different instance of AR(1) process with same $h$.

\begin{figure}[t]
\vspace*{-0.5cm}
	\centering
	\subfigure[]{\includegraphics[width=9.0 cm,height=4.5cm]{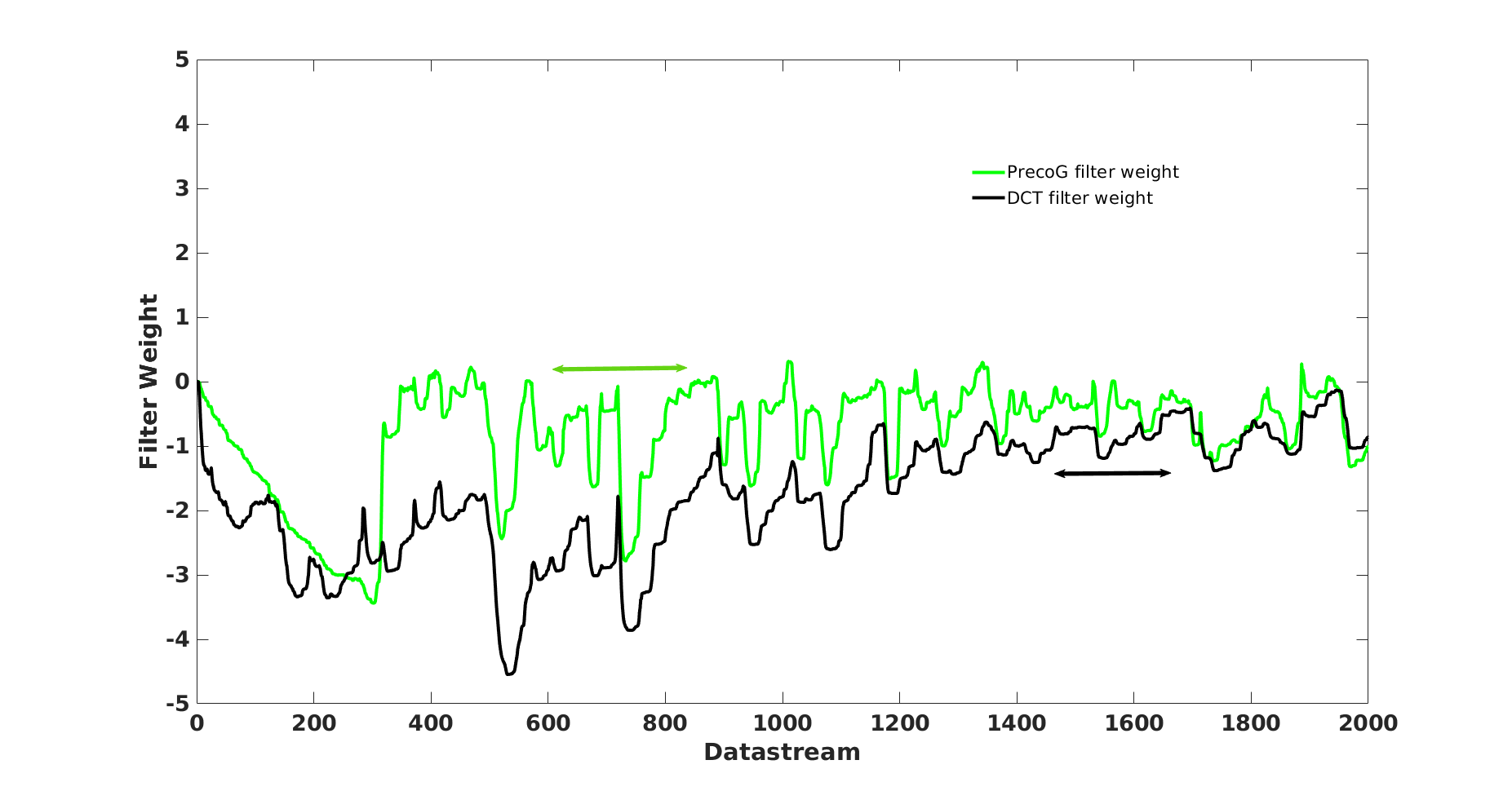}}
	\subfigure[]{\includegraphics[width=9.0 cm,height=2.5cm]{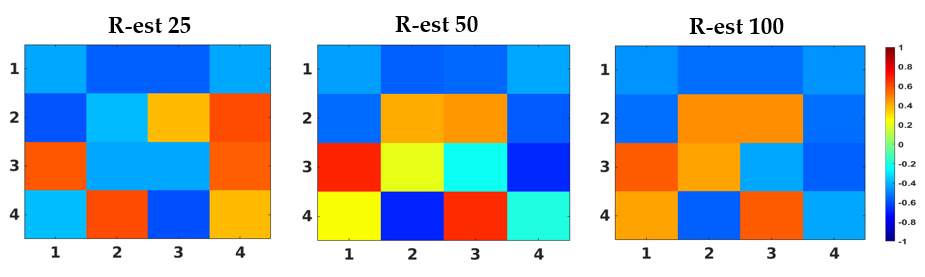}}
	\caption{(a) The plot shows convergence of a tap weight in cases of DCT and PrecoG, keeping the values of all the parameters same for both cases. The double-headed arrows mark the iteration where the weights started converging to the actual convolution weight.	(b) The PrecoG transformation matrices in case of an AR(1) process with $\rho=0.2$, estimated at different window length are shown. That because the data is comparatively decorrelated, PrecoG transformation contains pronounced inhibitory drives with small window length. AR(1) with higher correlated datastream demands longer window length as presented in Fig.~\ref{ar1fig}.  }
	\label{Ar1wgt}
\vspace*{-.5cm}
\end{figure}

A comparison between the DCT and PreoG in terms of the convergence of filter weights is given in Fig.~\ref{Ar1wgt}. The DCT is applied to the input data prior to resuming the LMS filter operation. PrecoG functions in two stages. At first, PrecoG computes the transformation using an initial data window (in this example, window length is $200$), during which the filter taps are updated without applying any transformation to the input. Once the transformation is obtained, it is applied to the rest of the data stream (TDLMS). It can be verified from Fig.\ref{ar1fig} that with DCT, the tap weight takes longer time to converge compared with PrecoG. The learning rate for LMS is static for both of these processes.

\begin{figure}[t]
\vspace*{0.5cm}
	\centering
	\subfigure[]{\includegraphics[width=9.0 cm,height=7.5cm]{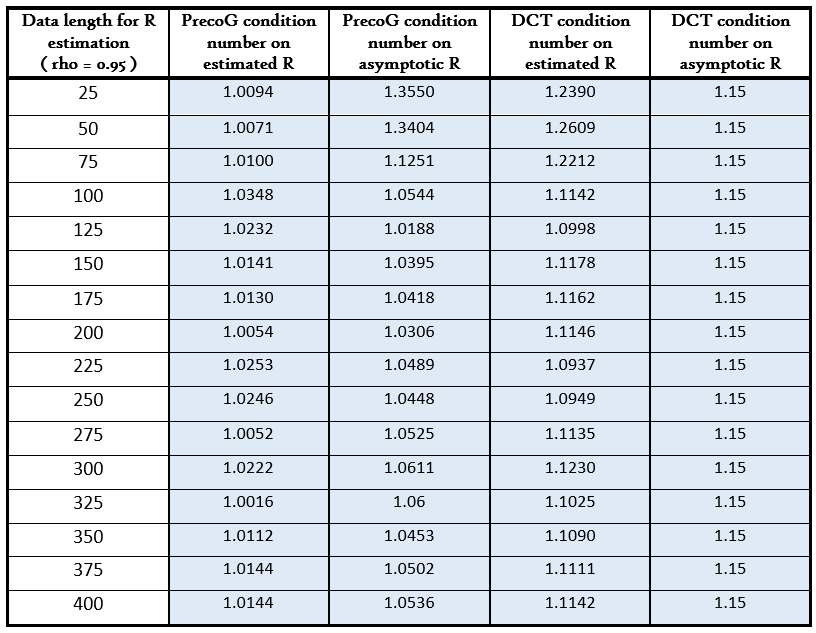}}
	\caption{\textbf{(Table 1)}The table lists the condition numbers that are obtained by the DCT and PrecoG transformations on AR(1) data with $\rho=0.95$. With different initial window lengths, the estimated autocorrelation matrices vary. The second and the fourth columns show the condition numbers by applying PrecoG and DCT on the estimated $R$ respectively. Using simple derivations, the asymptotic $R$ is given in eq.~\ref{acr}. The effect of PrecoG and the DCT are given in the third and fifth columns respectively. This table shows that with high signal correlation factor, such as $\rho=0.95$, we need at least $75$ length initial window to estimate $R$, when the condition number by PrecoG drops just below the DCT when applied on the asymptotic $R$.  }
	\label{Ar1table1}
\vspace*{-.5cm}
\end{figure}
\begin{figure}[t]
	\centering
	\subfigure[]{\includegraphics[width=9.0 cm,height=6.5cm]{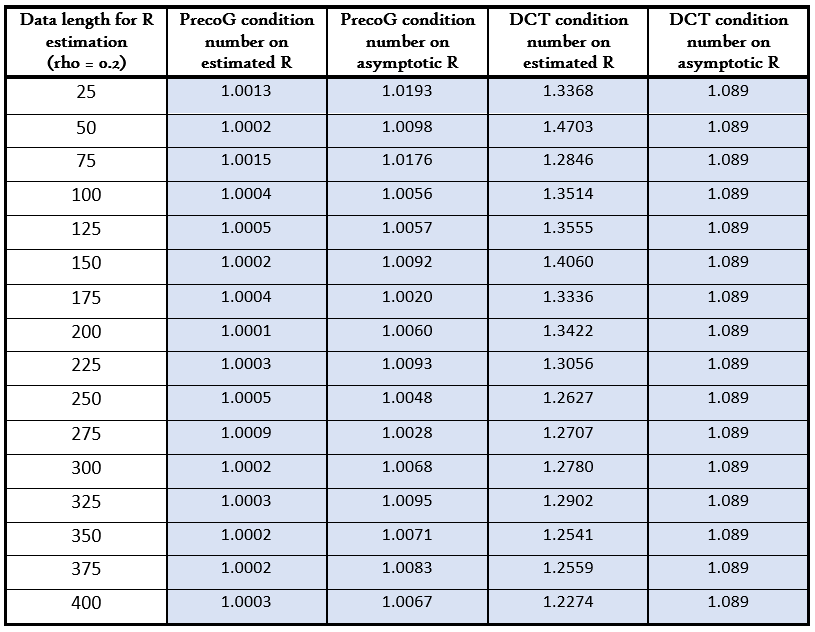}}	
	\caption{ \textbf{(Table 2)} The table lists the condition numbers that are obtained by the DCT and PrecoG transformations on AR(1) data with $\rho=0.2$. This table shows that with small signal correlation factor, such as $\rho=0.2$, we do not need to have a longer window to estimate $R$. In our simulation, at the initial window length of $25$, the condition number by PrecoG drops just below the DCT when applied on the asymptotic $R$. This makes sense as the data does not have enough correlation among adjacent samples because of low correlation factor. So, any preconditioning matrix will take short time to further decorrelate the $R$. }
	\label{Ar1table2}
\vspace*{-.5cm}
\end{figure}

A relevant question is: how long would the initial data window be in order to compute an effective preconditioner? Table 1 in Fig.~\ref{Ar1table1} and Table 2 in Fig.~\ref{Ar1table2} will provide answer to that question. In each table, we have provided results on a set of attributes. First, DCT and PrecoG are tested how they perform on the estimated time-averaged R ($R_{est}$) and asymptotic R (see eq.~\ref{acr}). Let us take the first row of Table 1. With the initial data window of length $25$, we have obtained a transformation $U$. The DCT gives the condition number of $1.2390$ on $R_{est}$.   
It might appear that PrecoG performed well over the DCT if the initial $25$ data samples are considered. However, the third column reveals that with $U$ (PrecoG) at hand, it yields the condition number of $1.3550$ when the asymptotic R is considered and it is inferior to the DCT ($1.15$). This due to the fact that the $R_{est}$ is not a robust approximation of the asymptotic $R$. 
It is not until the data window of length $75$, we can observe that PrecoG ($1.1251$) starts performing better than DCT ($1.15$) on the asymptotic R. 

Table 1 and 2 are the results on AR(1) process at two  different signal correlation factors $0.95$ and $0.2$ respectively. One can see that PrecoG starts performing better than DCT with the data window length $25$ in case of $\rho=0.2$. This observation is harmonious with the theory of TDLMS. AR(1) with $\rho=0.95$ is a highly correlated datastream. Therefore, a longer data window is required to estimate $R$ in order to decorrelate the data. Fig.~\ref{Ar1wgt}(b) shows the transformation matrices at three different initial length of data window for AR(1) process with $\rho=0.2$. When compared to Fig.~\ref{ar1fig}(a) (AR(1) with $\rho=0.95$), it is evident that
with smaller size data window, PrecoG contains a large number of inhibitory drives when the signal correlation factor is relatively low.

\subsection{$2^{nd}$ order autoregressive process}
\begin{figure}[t]
\vspace*{-0.5cm}
	\centering
	\subfigure[]{\includegraphics[width=8.0 cm,height=3.5cm]{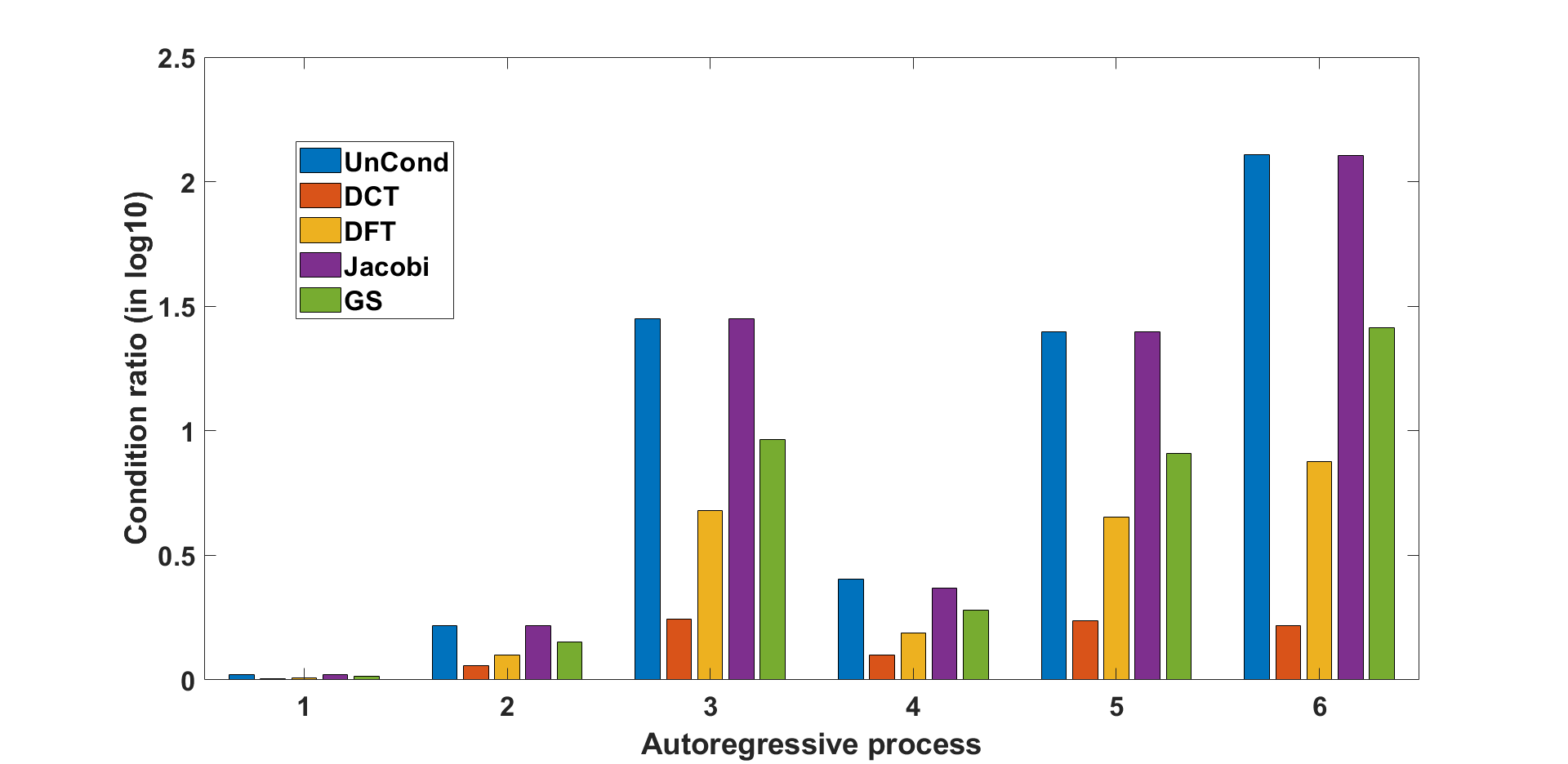}}
	\subfigure[]{\includegraphics[width=8.0 cm,height=3.5cm]{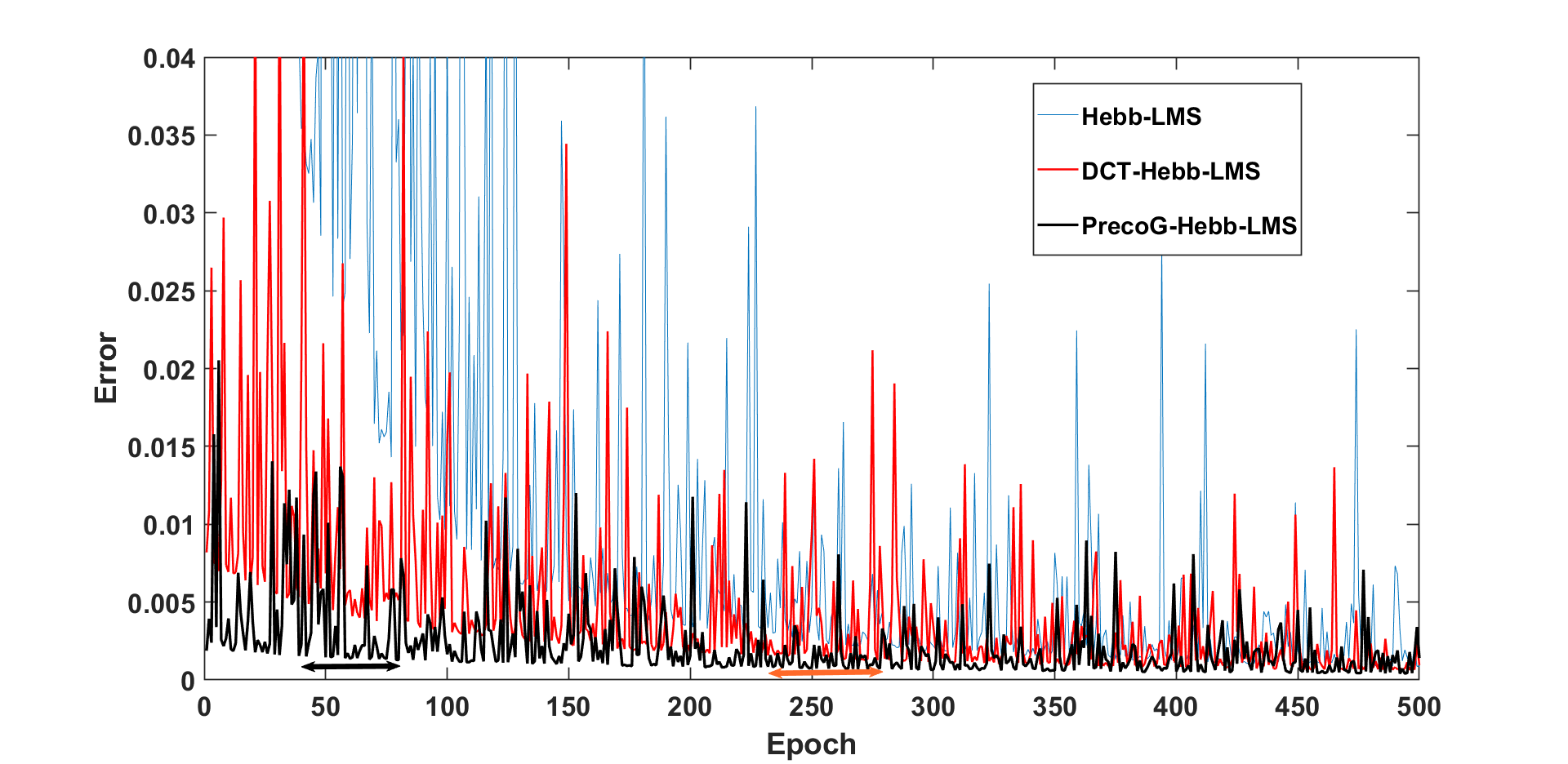}}
	\caption{(a) The plot shows comparative performances of DCT, DFT, Jacobi and GS with respect to PrecoG (condition ratio in log10) on six AR(2) datastreams with $(\rho_1,\rho_2)$ as  $(0.15, 0.1)$, $(0.75, 0.7)$, $(0.25,0.01)$, $(0.75, 0.1)$, $(0.9, 0.01)$ and $(0.99,0.7)$. (b) The error profile over epochs in case of Hebb-LMS, and the effect of PrecoG and the DCT on the Hebb-LMS process. With PrecoG, the error is reduced significantly within $60$ epochs.     }
	\label{Hebb-LMS}
\vspace*{-.5cm}
\end{figure}

In Fig.~\ref{Hebb-LMS}(a), we present the condition ratios computed by applying the algorithms on the autocorrelation matrices of eight $2^{nd}$ order autoregressive process with parameters $(\rho_1, \rho_2)$~\cite{zhao2009stability}.

The input autocorrelation matrix $R_N$ of such process is given by $R_N~=~c_1R_N(\rho_1)+c_2R_N(\rho_2)$. $R_N(\rho_1)$ and $R_N(\rho_2)$ are two Toeplitz matrices, similar to $R_N$ of $1^{st}$ order Markov process.
$c_1$ and $c_2$ are constants and are given by $c_1 = \frac{\rho_1(1-\rho_2^2)}{(\rho_1-\rho_2)(1+\rho_1\rho_2)}$, $c_2 = \frac{-\rho_2(1-\rho_1^2)}{(\rho_1-\rho_2)(1+\rho_1\rho_2)}$.
 As shown in Fig.\ref{Hebb-LMS}(a), PrecoG outperforms DCT in all the above cases, implying that it has better decorrelating ability than the DCT. 

\begin{figure}[t]
	\centering
	\subfigure[]{\includegraphics[width=8.0 cm,height=3.5cm]{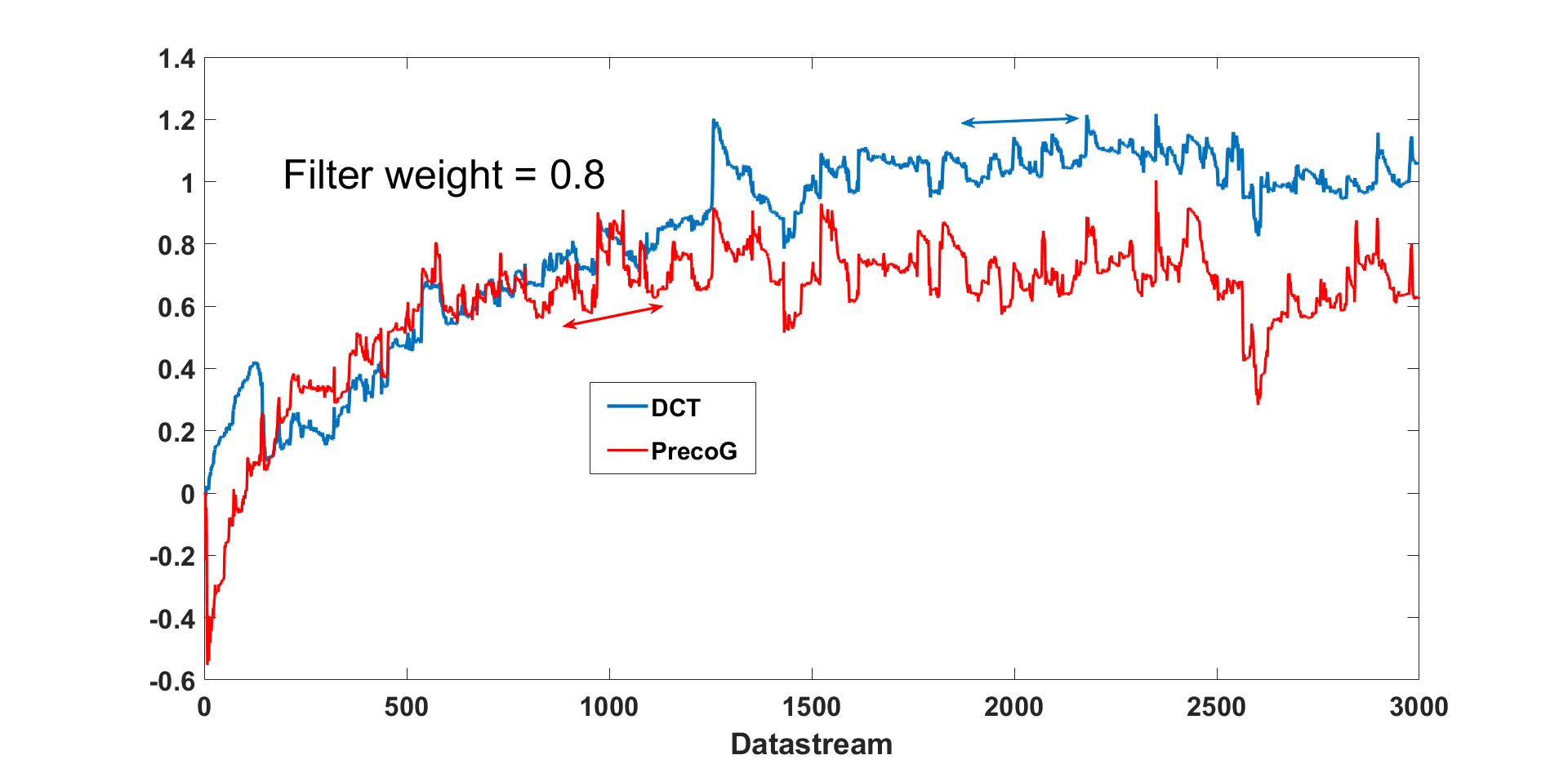}}	\subfigure[]{\includegraphics[width=8.0 cm,height=3.5cm]{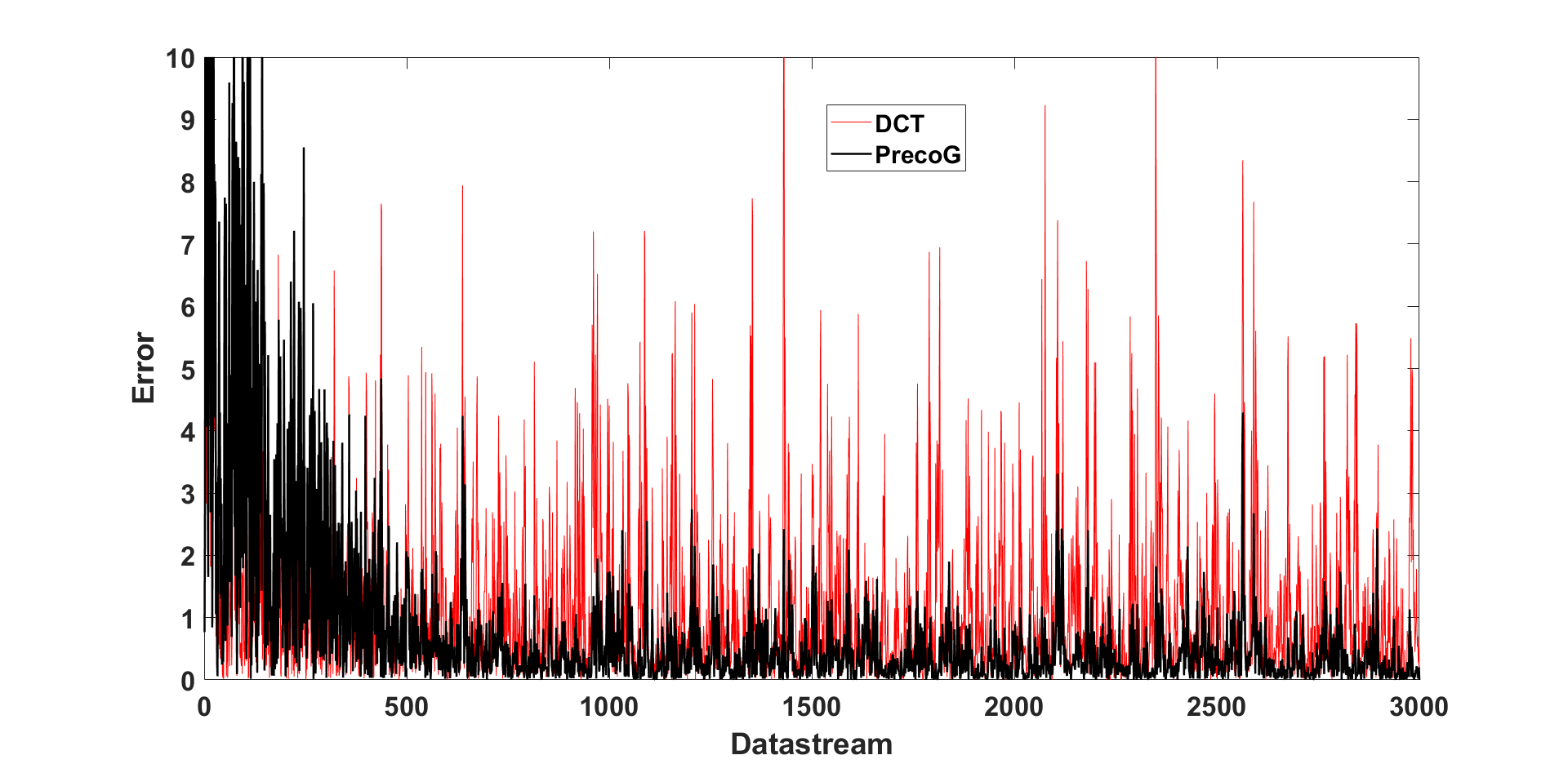}}
	\caption{(a) The plot shows the convergence of a filter weight in cases of the DCT and PrecoG when applied on an AR(2) process with $\rho_1=0.6$ and $\rho_2=0.9$. The double-headed arrows indicate the iteration interval where the tap weight starts converging. (b) The plot shows the error as the datastream progressess. With PrecoG, the error is rapidly diminished. With DCT, the error is reduced at first and then sustains. The parameters in both the experiments are kept same.    ) }
	\label{Ar2process}
\vspace*{-.2cm}
\end{figure}

Next, we apply PrecoG to a simulated $3200$-length data sample from an AR(2) process with $\rho_1=0.6$ and $\rho_2=0.9$.The AR(2) data power is set to unity.
We assume that there are three filter taps. The convolution filter $h = [1~0.8~-3]$ to generate the desired output. The convolved response is additively corrupted with white Gaussian noise with normalized signal power of $0.01$. 
For each data $x$, we set LMS learning step as $0.001$, and the power normalization factor $beta$ as $0.85$. The length of initial data window to estimate the autocorrelation matrix is set as $200$. Fig.~\ref{Ar2process}(a) shows the behavior of a filter tap in the cases of DCT-LMS and PrecoG-LMS where each datasample is processed. The tap weight  
attains convergence faster in PrecoG accompanied by an accelerated reduction in error (Fig.~\ref{Ar2process}(b)) when compared with the DCT.

\subsection{Hebbian learning and Hebb-LMS algorithm}
In 1949, Donald Hebb~\cite{hebb1949organization} postulated that concurrent synaptic changes occur as a function of pre and post synaptic activity, which was elegantly stated as "neurons that are wired together fire together." This classical remark of Hebb attempted to explain the plausible role of synaptic changes in learning and memory. Hebbian and LMS were predominantly regarded as two distinct forms of learning. Hebbian form of learning is unsupervised in nature, whereas LMS is primarily supervised. However, Hebbian rule, when translated into computational filters, leads to instability. Neuroscience researchers explained that neurons and the network collectively maintain stability by scaling the data at the input synapses (homeostatic plasticity)~\cite{turrigiano2004homeostatic,swanwick2006activity}. From the computational point of view, a parallel of this Hebbian learning and synaptic scaling with our work is given in \textbf{Appendix} (Hebb-LMS algorithm). In 2019, Widrow proposed unsupervised Hebb-LMS algorithm that contains an analogue of homeostatic plasticity while using a sigmoid. PrecoG is found to show its efficacy there.

The relevance of this Hebbian learning and homeostatic plasticity (synaptic scaling) in our work can be speculated from a computational point of view.   In supervised settings, the error at the output corresponding to an input data is backpropagated and eventually updates the weights. A data vector is an array of real numbers. The positive entries in a vector collectively constitute an excitatory drive and the negative entries constitute an inhibitory drive to the postsynaptic neuron, where the drives are summed. An ideal transformation (TDLMS) should have the property that it can regulate the magnitude of both the drives. A suitable transformation is expected to assess the balance of excitatory and inhibitory drives and modulate the input to achieve decorrelation. We resort to the input autocorrelation matrix that may provide such assessment. It is to remind the reader that input data is transformed prior to passing to the filter. Therefore, the transformation `scales' the data for the LMS filter.  So, it acts as a regularizer for the adaptation of filter weights. If this philosophy is true, PrecoG should also work in the unsupervised settings. In 2019, Widrow proposed unsupervised Hebb-LMS algorithm that contains an analogue of homeostatic plasticity while using a sigmoid. PrecoG is found to show its efficacy there.

\begin{figure}[t]
\vspace*{0.5cm}
	\centering
	\subfigure[]{\includegraphics[width=8.0 cm,height=3.5cm]{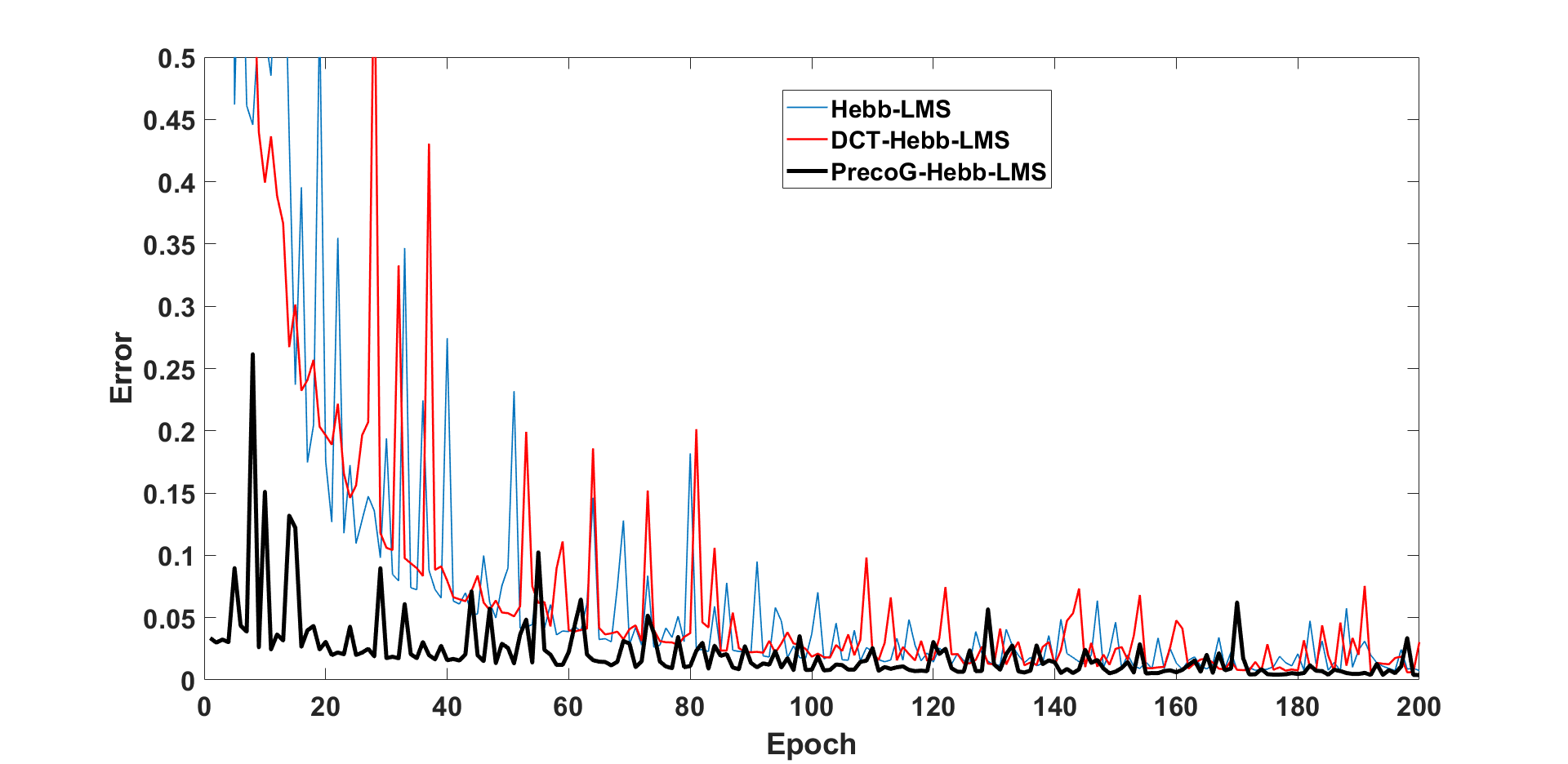}}
	\caption{(a) The error profile over epochs in case of Hebb-LMS, and the effect of PrecoG and the DCT on the Hebb-LMS process on an AR(2) data. The data is shuffled at each epoch prior to passing to the filter. $R$ is estimated from the shuffled in the first epoch. With PrecoG, the error is reduced significantly within $30$ epochs.      }
	\label{Hebb-LMS-2}
\vspace*{-.5cm}
\end{figure}

The experimental setting considers a $3200$ length data generated by an AR(1) process with a Gaussian noise with zero mean and a variance of $3$. We assume that there are four filter taps. For each data $x$, the desired output is $Sigmoid(W^{T}x)$. We set the $slpha$ of the sigmoid as $0.5$, $gamma$ as $0.5$, the LMS learning rate as $0.001$, and power normalization factor $beta$ as $0.85$. The length of initial data window to estimate the autocorrelation matrix is set as $100$. We run $500$ epochs to inspect the behavior of error. In each epoch, weights are continuously updated after each data sample is passed and the datastream is randomly shuffled. We take the Euclidean norm of the $3200$ errors in each epoch to plot. We keep this setting fixed for all Hebb-LMS, DCT-Hebb-LMS and PrecoG-Hebb-LMS on the data. Similar conclusion can be drawn when PrecoG and DCT are tested on an AR(2) process with a Gaussian noise with zero mean and variance $5$ (fig.~\ref{Hebb-LMS-2})

\section{Discussion}
\vspace*{-0.1cm}
LMS filters play pivotal roles in several applications, and convergence of the algorithm in terms of the filter weights, which are updated at each iteration, poses critical challenges to the quality of performance of LMS filters. TDLMS offers a solution to overcome the convergence issue by transforming the data prior to channeling it to the LMS filter. Conventional transforms serve as a palette of such transformation. However, the convergence of LMS is mediated by the shape of the real-life data manifold. If the shape of the data manifold is known \textit{a priori}, only then we can robustly evaluate the efficacy of the transformations. This issue was largely overlooked. PrecoG provides answer to that problem by presenting an optimization framework that intelligently encodes data manifold into the transformation matrix.   

Apart from symmetry by definition in case of real valued data, autocorrelation matrices containing distinct structures, such as diagonally dominant, Toeplitz, Hankel and circulant appear in special circumstances. For example, the autocorrelation matrix of $1^{st}$ order Markov process possesses Toeplitz structure. If the autocorrelation at two time points depends only on the time difference, then the autocorrelation matrix becomes Toeplitz.
However, these assumptions are scarcely valid in case of data with dynamic changes over time. PrecoG evaluates the data manifold after each interval and accordingly finds the unitary preconditioner. Users can set a time interval after which the preconditioning matrix will be computed periodically to incorporate dynamic changes into the matrix. Nevertheless, this flexibility comes at the price of minor additional computational overhead. To assess the computation time, we need to look at eq.\ref{finleqq} and Appendix. 

PrecoG is shown to significantly outperform the conventional transformation matrices that are considered in this paper on autoregressive datasets and different linear systems of equations. PrecoG has been shown ( in Table 1 and Table 2 ) to have improved performance when the autocorrelation matrix of the process is estimated, making PrecoG amenable to be deployed in on-line scenarios. For autoregressive datasets, we also examined the variation of the initial window length with the signal correlation factor. It was shown to be effective in unsupervised settings also, when we investigated PrecoG's performance in the Hebbian-LMS settings.

\vspace*{-0.2cm}
\section{Conclusion}
\vspace*{-0.3cm}
In this work, we present a method to obtain a unitary split preconditioner by utilizing nonconvex optimization, graph theory and first-order perturbation theory. We demonstrate the efficacy of our approach over prevalent state-of-the-art techniques on Markov datasets and linear systems of equations. We inspect PrecoG in supervised and unsupervised settings. 
As a future endeavor, we will attempt to exploit the signal structure and embed this structure into the optimization framework by including a set of constraints. In continuation, we will try to extend our approach to solve a sparse underdetermined linear system of equations in order to implement dictionary learning. In addition, we will inspect the behavior of TDLMS filters by assigning the characteristics of computational neurons on the LMS taps.


\vspace*{0.2cm}
\section*{Appendices}
\vspace*{-0.2cm}
\label{appn}
\noindent\textbf{LMS and TDLMS algorithms}\\
Let $X$ be a real-valued data that is channeled to a filter with $N$ tap delays, $W(n) = [w_0,w_1,...w(N-1)]$. The data vector at time $t=N$ at the taps can be written as $x(N) = [x_{N-1},x_{N-2},...,x_0]$. The filter output is given by $y(n) = W(n)^Tx(n)$. Let the desired response be $d(n)$. Then, the error $e(n) = d(n)-y(n)=d(n)-W(n)^Tx(n)$. 

The LMS algorithm is formulated as follows.
\begin{eqnarray}
\label{0a}
w_i(n+1) &=& w_i(n) - \mu\frac{\partial e^2(n)}{\partial w_i(n)}\nonumber\\
&=&  w_i(n) - 2\mu e(n)\frac{\partial[d(n) - W(n)^Tx(n)]}{\partial w_i(n)}\nonumber\\
&=& w_i(n) + 2\mu e(n)x(n-i)
\end{eqnarray}
Assume that W(n) is independent of X(n) (because at the convergence, the values of W(n) will be stationary but X(n) will keep changing). Taking expectation on both sides of the above expression, we get
\begin{eqnarray}
\label{1a}
E(W(n+1)) &=& E(W(n))+2\mu E\big(e(n)x(n)\big)\nonumber\\
 &=& E(W(n))+2\mu E\big(x(n)(d(n)\nonumber\\
 &&-x(n)^TW(n))\big)\nonumber\\
 &=& E(W(n))+2\mu E(x(n)d(n)) \nonumber\\
 &&- 2\mu E(x(n)x(n)^T)E(W(n))\nonumber\\
 &=& E(W(n))+2\mu R_{dx} -2\mu R_{xx}E(W(n))\nonumber\\
 &=& \big(1-2\mu R_{xx}\big)E(W(n)) + 2\mu R_{dx}
\end{eqnarray}
Let us consider the term $E(W(n))\big(1-2\mu R_{xx}\big)$. let us assume $R_{xx} = R$ and $R_{xx} = U\Lambda U^T$. Therefore,
\begin{eqnarray}
E(W(n))\big(1-2\mu R_{xx}\big) &=& U\big(I - 2\mu\Lambda\big)U^TE(W(n))\nonumber\\
&=& U\big(I - 2\mu\Lambda\big)^nU^TE(W(0))
\end{eqnarray}
In order to converge $E(W(n+1))$ to a stable value $2\mu R_{dx}$ in eq.~(\ref{1a}), $\big(I - 2\mu\Lambda\big)^n$ must converge to zero at $n\to \infty$. This is possible only when $(1-2\mu\lambda_i)\to 0 \forall i$. So, the convergence will be quickly achieved if $\frac{\lambda_{max}}{\lambda_{min}}\to 1$. 

In TDLMS, let the transformation be $T\in\mathcal{R}^{N\times N}$ and the transformed data vector is $x^{'}(N) = Tx(N)$. The following step is the normalization of power. 
\begin{eqnarray}
v_i(n) &=& \frac{x_i^{'}(n)}{\sqrt{P_i(n)+eps}}; eps=\text{small number}\nonumber\\
P_i(n) &=& \beta P_i(n-1)+(1-\beta)(x_i^{'})^2(n)
\end{eqnarray}
Next, $v(n)$ is channeled to the LMS filter. the expression can be obtained simply by replacing $x(n-i)$ with $v(n-i)$ in eq.~(\ref{0a}).

\vspace{1cm}
\noindent\textbf{Hebb-LMS algorithm}\\
In 1949, Donald Hebb~\cite{hebb1949organization} postulated that concurrent synaptic changes occur as a function of pre and post synaptic activity, which was elegantly stated as "neurons that are wired together fire together". This classical remark of Hebb attempted to explain the plausible role of synaptic changes in learning and memory. However, it was just a conjecture at that time. Twenty years later, in a landmark paper, Bliss and Lomo~\cite{bliss1973long} showed the existence of long-term potentiation (LTP) with the help of applying a train of brief, high frequency stimulation, known as tetanus, to the Hippocampus. This discovery led to a quest to mine the properties of LTP in the context of excitatory synaptic transmission~\cite{mcnaughton1978synaptic,levy1979synapses}.

So, Hebbian and LMS were predominantly regarded as two distinct forms of learning. Hebbian form of learning is unsupervised in nature, whereas LMS is primarily supervised. However, when Hebb's rule was applied to a linear filter to learn patterns in data by adaptively changing the weights, it posed a problem. If the concurrent pre and post synaptic activities strengthen over time, there is nothing to scale the activity down. Researchers attempted to introduce a `forgetting' term inside the Hebb's rule.
In 1982, Oja~\cite{oja1982simplified} introduced a modification to stabilize the computational analogue of Hebb rule. In effect, the linear filter was able to learn the principal components of the input data. However, such modifications are \textit{ad hoc} and, therefore, difficult to interpret. 

In neuroscience, there exists a trenchant question: how does this intricate web of neurons, that is constantly undergoing physical changes maintains stability (homeostasis)? Researchers explained that neurons and the network collectively maintain the stability (for example, by regulating the average firing rate of a set of neurons)~\cite{marder2003current,turrigiano2004homeostatic}. One such mechanism that gives stability is called synaptic scaling~\cite{turrigiano2004homeostatic,swanwick2006activity}. 

Widrow~\cite{widrow2019nature} proposed an algorithm that integrates Hebbian and LMS learning paradigms and it contains two equilibrium states for excitatory and inhibitory drives. The mathematical expression is given by 
\begin{eqnarray}
W(n+1) &=& W(n) + 2\mu e(n)x(n)\nonumber\\
e(n) &=& SGM(W(n)^Tx(n)) - \gamma W(n)^Tx(n)\nonumber\\
y(n) &=& \begin{cases}
SGM(W(n)^Tx(n)) & if~SGM(W(n)^Tx(n))>0\\
0 & \text{o.w.}
\end{cases}\nonumber\\
\end{eqnarray}
Here $SGM$ is the polar sigmoid function. 
This algorithm is unsupervised because the target response of x(n) is $SGM(W(n)^Tx(n))$. It contains two stable equilibrium points $1$ (excitatory) and $-1$ (inhibitory) on the SGM curve.

\vspace{1cm}
\noindent\textbf{Evaluation of $\big[\frac{\partial E(p)}{\partial U_{N}}\big]$}\\
 \vspace{-0.2cm}
Let, $M(\epsilon) = ||U^TRU - (1+\epsilon)U^TRU\circ\mathcal{I}||_F^2$. Then,

\begin{eqnarray}
\label{eqAp1}
M(\epsilon) &=& Tr\Big(\{U^TRU - (1+\epsilon)U^TRU\circ\mathcal{I}\}\nonumber\\
	&&\{U^TRU - (1+\epsilon)U^TRU\circ\mathcal{I}\}^T \Big).
\end{eqnarray} 
Eq.~(\ref{eqAp1}) on expansion gives
\begin{eqnarray}
\label{eqAp2}
M(\epsilon) &=& Tr\Big(U^TR^2U - 2(1+\epsilon)(U^TRU\circ\mathcal{I})(\nonumber\\
&&U^TRU) + (1+\epsilon)^2(U^TRU\circ\mathcal{I})^2 \Big).
\end{eqnarray}
Eq.~(\ref{eqAp2}) is obtained using the fact that $UU^T = \mathcal{I}$. By performing the partial derivative,
\begin{eqnarray}
\frac{\partial M(\epsilon)}{\partial U} &=& \frac{\partial Tr(U^TR^2U)}{\partial U} -2(1+\epsilon)\frac{\partial }{\partial U}Tr\{(U^TRU\circ\mathcal{I})\nonumber\\
						&& (U^TRU)\} + (1+\epsilon)^2\frac{\partial }{\partial U}Tr(U^TRU\circ\mathcal{I})^2.\nonumber\\
						&=& 2R^2U - 4(1+\epsilon)RU + (1+\epsilon)^2(U^TRU\circ\mathcal{I})RU\nonumber
\end{eqnarray}
Next, $\frac{\partial E(p)}{\partial U}$ is computed using $\frac{\partial M}{\partial U}$ as $\frac{\partial E(p)}{\partial U} = \frac{\partial M(+\epsilon_1)}{\partial U} + \frac{\partial M(-\epsilon_2)}{\partial U}$.
\begin{eqnarray}
\frac{\partial E(p)}{\partial U_{n}} &=& 2\Big[2R_n - 2(2-\epsilon_2-\epsilon_2)\mathcal{I} - \{(1+\epsilon_1)^2 + \nonumber\\
&&(1-\epsilon_2)^2\}U_n^TR_nU_n\circ\mathcal{I}\Big]R_nU_n.\nonumber\\
\end{eqnarray}
\vspace*{-0.3cm}

\begin{eqnarray}
\frac{\partial M(\epsilon)}{\partial U} &=& \frac{\partial Tr(U^TR^2U)}{\partial U} -2(1+\epsilon)\frac{\partial }{\partial U}Tr\{(U^TRU\circ\mathcal{I})\nonumber\\
						&& (U^TRU)\} + (1+\epsilon)^2\frac{\partial }{\partial U}Tr(U^TRU\circ\mathcal{I})^2.\nonumber\\
						&=& 2R^2U - \big[8(1+\epsilon)+4(1+\epsilon)^2\big](U^TRU\circ\mathcal{I})RU\nonumber
\end{eqnarray}

\vspace*{1cm}
\noindent\textbf{Proof of eq.~(15)}\\
Let us assume $A_0$ is a real-valued, finite-dimensional ($\in\mathcal{R}^{N}$), positive semi-definite matrix with eigenvectors $U_0 = [\boldsymbol{u}_1, \boldsymbol{u}_2,...,\boldsymbol{u}_N]$ and the corresponding eigenvalues $\lambda_0 = [\lambda_1,\lambda_2,...\lambda_N]$ (arranged in the decreasing order). In our case, $A_0 = L$, the graph Laplacian matrix. By definition $\lambda_N = 0$. 

Let us also assume that $A$, $U$ and $\lambda$ are real-valued functions of a continuous parameter $\tau$ and they are analytic in the neighborhood of $\tau_0$ such that $A(\tau_0) = A_0$, $U(\tau_0)= U_0$ and $\lambda(\tau_0)=\lambda_0$.
\begin{eqnarray}
\label{auxx0}
A(\tau)\boldsymbol{u}_i(\tau) &=& \lambda_i(\tau)\boldsymbol{u}_i(\tau)\nonumber\\
\dot{A}(\tau)\boldsymbol{u}_i(\tau) + A(\tau)\dot{\boldsymbol{u}}_i  &=& \dot{\lambda}_i(\tau)\boldsymbol{u}_i+\lambda_i\dot{\boldsymbol{u}}_i 
\end{eqnarray}
Now, because of the facts that $\boldsymbol{u}_i$ is analytic and $||\boldsymbol{u}_i|| = 1$ (orthonormal), $\dot{\boldsymbol{u}}_i \bot \boldsymbol{u}_i$. 
Taking inner product with $\boldsymbol{u}_i$ on both sides and plugging $\big<\dot{\boldsymbol{u}}_i,\boldsymbol{u}_i\big>=0$, it can be found that
\begin{eqnarray}
\label{auxx1}
\big<\dot{A}\boldsymbol{u}_i,\boldsymbol{u}_i\big> + \big<A\dot{\boldsymbol{u}}_i,\boldsymbol{u}_i\big> &=& \big<\dot{\lambda}\boldsymbol{u}_i,\boldsymbol{u}_i\big>\nonumber\\
\big<\dot{A}\boldsymbol{u}_i,\boldsymbol{u}_i\big> &=& \big<\dot{\lambda}\boldsymbol{u}_i,\boldsymbol{u}_i\big>.
\end{eqnarray}
Here, $\big<A\dot{\boldsymbol{u}}_i,\boldsymbol{u}_i\big> = \big<\dot{\boldsymbol{u}}_i,A\boldsymbol{u}_i\big>$ (because $A$ is symmetric (self-adjoint)). Using $A\boldsymbol{u}_i = \lambda_i\boldsymbol{u}_i$, we get $\big<\dot{\boldsymbol{u}}_i,A\boldsymbol{u}_i\big>$ = $\big<\dot{\boldsymbol{u}}_i,\lambda_i\boldsymbol{u}_i\big>$= $\lambda_i\big<\dot{\boldsymbol{u}}_i,\boldsymbol{u}_i\big> = 0$.
From eq.~\ref{auxx1}, we obtain the expression for $\dot{\lambda}_i$ as
$\dot{\lambda}_i = \big<\dot{A}\boldsymbol{u}_i,\boldsymbol{u}_i\big>$. 

Taking the inner product of eq.~\ref{auxx0} with $\boldsymbol{u}_j;~j\neq i$ and using the fact that $\boldsymbol{u}_i\bot\boldsymbol{u}_j$ we obtain
\begin{eqnarray}
\label{auxx00}
\big<\dot{A}\boldsymbol{u}_i, \boldsymbol{u}_j\big> + \big<A\dot{\boldsymbol{u}}_i,\boldsymbol{u}_j\big> &=& \lambda_i\big<\dot{\boldsymbol{u}}_i,\boldsymbol{u}_j\big>\nonumber\\
\big<\dot{A}\boldsymbol{u}_i, \boldsymbol{u}_j\big> + \big<\dot{\boldsymbol{u}}_i,A\boldsymbol{u}_j\big> &=& \lambda_i\big<\dot{\boldsymbol{u}}_i,\boldsymbol{u}_j\big>\nonumber\\
\big<\dot{A}\boldsymbol{u}_i, \boldsymbol{u}_j\big> + \big<\dot{\boldsymbol{u}}_i,\lambda_j\boldsymbol{u}_j\big> &=& \lambda_i\big<\dot{\boldsymbol{u}}_i,\boldsymbol{u}_j\big>\nonumber\\
\big<\dot{A}\boldsymbol{u}_i, \boldsymbol{u}_j\big> &=& (\lambda_i-\lambda_j)\big<\dot{\boldsymbol{u}}_i,\boldsymbol{u}_j\big>;~\lambda_i\neq\lambda_j\nonumber\\
\big<\dot{\boldsymbol{u}}_i,\boldsymbol{u}_j\big> &=& \frac{1}{(\lambda_i-\lambda_j)}\big<\dot{A}\boldsymbol{u}_i, \boldsymbol{u}_j\big>;~\lambda_i\neq\lambda_j \nonumber\\
\end{eqnarray}

Let us consider $\tau = a_{mn}$, where $a_{mn}$ is an entry of $A$. 

\begin{eqnarray}
\label{auxx3}
\frac{\partial \boldsymbol{u}_i}{\partial a_{mn}} = \sum_{p \neq i}\frac{1}{(\lambda_i-\lambda_p)}\big<\frac{\partial A}{\partial a_{mn}}\boldsymbol{u}_i,\boldsymbol{u}_p\big>\boldsymbol{u}_p;~ \lambda_i\neq\lambda_p.
\end{eqnarray}
It can be seen that $\dot{\boldsymbol{u}}_i$ is spanned by the eigenvectors of $A$, when all the eigenvalues are numerically different. Due to the fact that $A$ is symmetric, $\frac{\partial A}{\partial a_{mn}}$ is also symmetric. Therefore, eq.~\ref{auxx3} can be rearranged as
\begin{eqnarray}
\label{auxx4}
\frac{\partial \boldsymbol{u}_i}{\partial a_{mn}} = \sum_{p \neq i}\frac{1}{(\lambda_i-\lambda_p)}\big<\boldsymbol{u}_i,\frac{\partial A}{\partial a_{mn}}\boldsymbol{u}_p\big>\boldsymbol{u}_p;~ \lambda_i\neq\lambda_p.\nonumber\\
\end{eqnarray}
The above expression is valid for $\lambda_i\neq\lambda_j$. We can use the fourth step of eq.~\ref{auxx00} to derive the expression of $\dot{\boldsymbol{u}}_i$ in case of $\lambda_i=\lambda_j$ (multiplicity of eigenvalues). 

\begin{eqnarray}
\label{auxx11}
\big<\dot{A}\boldsymbol{u}_i, \boldsymbol{u}_j\big> &=& 0;~\lambda_i=\lambda_j.\nonumber\\
\big<\dot{A}\boldsymbol{u}_i, \boldsymbol{u}_j\big> &=& \lambda_i\big<\dot{\boldsymbol{u}}_i,\boldsymbol{u}_i\big>;~\dot{\boldsymbol{u}}_i\bot\boldsymbol{u}_i.\nonumber\\
\big<\boldsymbol{u}_i, \dot{A}\boldsymbol{u}_j\big> &=& \lambda_i\big<\dot{\boldsymbol{u}}_i,\boldsymbol{u}_i\big>;~\dot{A}~is~symmetric.\nonumber\\
\big<\dot{A}\boldsymbol{u}_j, \boldsymbol{u}_i\big> &=& \lambda_i\big<\dot{\boldsymbol{u}}_i,\boldsymbol{u}_i\big>.\nonumber\\
\dot{\boldsymbol{u}}_i &=& \begin{cases}
\frac{1}{\lambda_i}\sum_{j\neq i}\big<\dot{A}\boldsymbol{u}_i,\boldsymbol{u}_j\big>\boldsymbol{u}_i; & \lambda_i=\lambda_j\neq 0\\
\sum_{j\neq i}\big<\dot{A}\boldsymbol{u}_i,\boldsymbol{u}_j\big>\boldsymbol{u}_i; &\lambda_i=\lambda_j= 0
\end{cases}
\end{eqnarray}

Combining eq.~\ref{auxx11} and~\ref{auxx4}, 
\begin{eqnarray}
\label{finalauxx}
\frac{\partial \boldsymbol{u}_i}{\partial a_{mn}} &=& \sum_{p \neq i}\frac{1-\delta(\lambda_i,\lambda_p)}{(\lambda_i-\lambda_p)}\big<\boldsymbol{u}_i,\frac{\partial A}{\partial a_{mn}}\boldsymbol{u}_p\big>\boldsymbol{u}_p\nonumber\\ 
&&+ \sum_{\substack{
q\neq i\\
\lambda_q\neq 0}}\frac{\delta(\lambda_i,\lambda_q)}{\lambda_i}\big<\boldsymbol{u}_i,\frac{\partial A}{\partial a_{mn}}\boldsymbol{u}_q\big>\boldsymbol{u}_i \nonumber\\
&& + \sum_{\substack{
q\neq i\\
\lambda_q= 0}}\delta(\lambda_i,\lambda_q)\big<\boldsymbol{u}_i,\frac{\partial A}{\partial a_{mn}}\boldsymbol{u}_q\big>\boldsymbol{u}_i
\end{eqnarray}
Here $\delta$ is the Kronecker delta function.

\vspace*{0.2cm}
\noindent\textbf{Proof of eq.(16)}\\
By definition,
\begin{eqnarray}
\frac{\partial \boldsymbol{u}_k}{\partial w_j} = \sum_{m,n}\frac{\partial \boldsymbol{u}_k}{\partial L_{mn}}\frac{\partial L_{mn}}{\partial w_j},
\end{eqnarray}
where $L_{mn}$ indicates the $(m,n)^{th}$ element of the symmetric matrix $L$. 
Let us first find the expression for $\lambda_i\neq\lambda_j$. 
Inserting eq.~\ref{auxx4} by replacing $A$ with $L$ to the above equation, we get
\begin{eqnarray}
\label{auxx5}
\frac{\partial \boldsymbol{u}_k}{\partial w_j} &=& \sum_{m,n}\Big[\sum_{k\neq q}\frac{1}{(\lambda_k-\lambda_q)}\big<\frac{\partial L}{\partial L_{mn}}\boldsymbol{u}_k,\boldsymbol{u}_q\big>\boldsymbol{u}_q\Big]\frac{\partial L_{mn}}{\partial w_j}, \nonumber\\
&=& \sum_{k\neq q}\Big(\frac{1}{\lambda_k-\lambda_q}\Big)\Big[\sum_{m,n}\big<\frac{\partial L}{\partial L_{mn}}\boldsymbol{u}_k,\boldsymbol{u}_q\big>\boldsymbol{u}_q\Big]\frac{\partial L_{mn}}{\partial w_j},\nonumber\\
&=& \sum_{k\neq q}\Big(\frac{1}{\lambda_k-\lambda_q}\Big)\Big[\sum_{m,n}\big<\boldsymbol{u}_k,\frac{\partial L}{\partial L_{mn}}\boldsymbol{u}_q\big>\boldsymbol{u}_q\Big]\frac{\partial L_{mn}}{\partial w_j},\nonumber\\
&=& \sum_{k\neq q}\Big(\frac{1}{\lambda_k-\lambda_q}\Big)\big<\boldsymbol{u}_k, \sum_{m,n}\Big(\frac{\partial L}{\partial L_{mn}}\frac{\partial L_{mn}}{\partial w_j}\Big)\boldsymbol{u}_q\big>\boldsymbol{u}_q\nonumber\\
&=& \sum_{k\neq q}\Big(\frac{1}{\lambda_k-\lambda_q}\Big)\big<\boldsymbol{u}_k, B\frac{\partial W}{\partial w_j}B^T\boldsymbol{u}_q\big>\boldsymbol{u}_q\nonumber\\
&=& \sum_{k\neq q}\big<\boldsymbol{u}_k, \frac{1}{(\lambda_k-\lambda_q)}B\frac{\partial W}{\partial w_j}B^T\boldsymbol{u}_q\big>\boldsymbol{u}_q
\end{eqnarray}
From eq.~\ref{auxx5}, it can be observed that the eigenvector $\boldsymbol{u}_q$ is projected by the operator $ \frac{1}{(\lambda_k-\lambda_q)}B\frac{\partial W}{\partial w_j}B^T$ prior to the inner product.\\
Applying the same definition the final expression can be found by using eq.~\ref{finalauxx}. 
\begin{eqnarray}
\label{finalxxx}
\frac{\partial \boldsymbol{u}_i}{\partial w_j} &=& \sum_{\substack{p \neq i \\
\lambda_p\neq\lambda_i}
}\frac{1-\delta(\lambda_i,\lambda_p)}{(\lambda_i-\lambda_p)}\big<\boldsymbol{u}_i,B\frac{\partial W}{\partial w_{j}}B^T\boldsymbol{u}_p\big>\boldsymbol{u}_p\nonumber\\
&&+ \sum_{\substack{q\neq i \\ 
\lambda_q=\lambda_i\neq 0}}
\frac{\delta(\lambda_i,\lambda_q)}{\lambda_i}\big<\boldsymbol{u}_i,B\frac{\partial W}{\partial w_{j}}B^T\boldsymbol{u}_q\big>\boldsymbol{u}_i.\nonumber\\
&& +\sum_{\substack{q\neq i \\ 
\lambda_q=\lambda_i= 0}}
\delta(\lambda_i,\lambda_q)\big<\boldsymbol{u}_i,B\frac{\partial W}{\partial w_{j}}B^T\boldsymbol{u}_q\big>\boldsymbol{u}_i.
\end{eqnarray}

\vspace*{-.2cm}
\noindent\textbf{Proof: $\frac{\partial L}{\partial u_{kl}} = U_N\Gamma J^{mn} + J^{nm}\Gamma U_N^T $} is non-invertible\\
By definition,  $J^{mn} = \delta_{mk}\delta_{np}$. Let $F = \Gamma J^{mn}$. Then 
\begin{eqnarray}
F(x,y) = 
\begin{cases}
\lambda_k & \quad \text{if x=k and y=p}\\
0 & \quad \text{o.w.}
\end{cases}
\end{eqnarray}
Therefore, 
\begin{eqnarray}
U_N\Gamma J^{mn}(x,y) = 
\begin{cases}
\lambda_ku_{xk} & \text{if y=p} \\
0 & \text{o.w.}
\end{cases}
\end{eqnarray}
So, $U_N\Gamma J^{mn}$ has one column (at $y=p$) that has a set of non-zero entries.  
It is evident that $H = (U_N\Gamma J^{mn})^T = J^{nm}\Gamma U_N^T$. Therefore, $\frac{\partial L}{\partial u_{kl}}$ has one row and one column of possibly non-zero entries by construction. Assume $m$ is that row index and the corresponding entry of $H$ is $[H_{m1}, H_{m2},...,H_{mN}]$. Assume that $q\in\{1,2,...,N\}$ is the column index, where $H$ receives entries from $U_N\Gamma J^{mn}$. One can find any two columns $i,j\in\{1,2,...,N\}-\{q\}$ such that the following transformation $C_i\longrightarrow \frac{C_i}{H_{mi}}$ and $C_j\longrightarrow\frac{C_j}{H_{mj}}$ would give two identical columns.

\vspace*{1cm}
\noindent\textbf{Complexity}\\

The update of weights $w_{i}$ requires the computation of three partial derivatives (see Appendix).
\noindent\textit{Using eq.~\ref{finalxxx}}\\
 Note that $B\frac{\partial W}{\partial w_{j}}B^T$ is fixed for $w_j$. By definition, $L = BWB^T$. So, $E=\frac{\partial L}{\partial w_j} = B\frac{\partial W}{\partial w_{j}}B^T$.  For a simple, connected graph, there is no self-loop. Let us assume $w_j = w_{kq}$, meaning that the edge between the $k^{th}$ and $q^{th}$ vertices has a weight $w_j$. $L$ is a linear function of $w_j$. Then, $E$ has exactly four non-zero entries - $E(k,k)=E(q,q) = 1$, $E(k,q)=E(q,k)=-1$. Therefore, $B\frac{\partial W}{\partial w_{j}}B^T\boldsymbol{u}_p$ requires constant computation. For example, if $w_j = w_{13}$, then $B\frac{\partial W}{\partial w_{j}}B^T\boldsymbol{u}_p$= $[u_{p1}-u_{p3}, 0, u_{p3}-u_{p1},0,0,..]^T$, having non-zer entries at the $1^{st}$ and $3^{rd}$ locations.
It implies that $\big<\boldsymbol{u}_i,B\frac{\partial W}{\partial w_{j}}B^T\boldsymbol{u}_q\big>$ requires constant computation. Now, each $\boldsymbol{u}_i$ has length $N$. For each $i$, $\frac{\partial \boldsymbol{u}_i}{\partial w_j}$ requires $\mathcal{O}(N)$ time to add for all $(N-1)$ vectors. So, the time complexity to compute $\frac{\partial U_N}{\partial w_j}$ is $\mathcal{O}(N^2)$, which is a significant improvement over the last scheme. The improvement is due to the non-existence of matrix inversion.    \\

It is to note that the constructions of preconditioners for solving linear systems by comparative methods such as, Jacobi~($\mathcal{O}(N^2)$), successive over-relaxation (SOR)~($\mathcal{O}(N^3)$), symmetric SOR~($\mathcal{O}(N^3)$), Gauss-Seidel~($\mathcal{O}(N^3)$) have faster associated run times compared to PrecoG. Here, complexity accounts for the inversion of each preconditioner matrix. However, the acceleration in the convergence of the LMS filter using PrecoG is also expected to partially compensate for the computational overload of PrecoG.


\bibliographystyle{unsrt}

\end{document}